\documentclass[journal, onecolumn, draft]{IEEEtran}
\IEEEoverridecommandlockouts

\usepackage{hyperref}
\usepackage{cite}
\usepackage{enumerate}
\usepackage{amsmath,amssymb,amsfonts}
\usepackage{graphicx, subcaption}
\usepackage{textcomp}
\usepackage{blindtext}
\usepackage[inline]{enumitem}
\usepackage{algorithm}
\usepackage{algpseudocode}
\usepackage{color}

\def \bgamma       {\boldsymbol{\gamma}}
\def \bGamma       {\boldsymbol{\Gamma}}

\def \bepsilon     {\boldsymbol{\epsilon}}

\def \blambda      {\boldsymbol{\lambda}}

\def \mbs          {\mathbf{s}}

\def \mby          {\mathbf{y}}

\def \mbA          {\mathbf{A}}

\def \complexC     {\mathbb{C}}

\def \naturalN     {\mathbb{N}}

\def \realR        {\mathbb{R}}

\def \calN         {\mathcal{N}}

\def \asin         {\sin^{-1}}

\def\BibTeX{{\rm B\kern-.05em{\sc i\kern-.025em b}\kern-.08em
		T\kern-.1667em\lower.7ex\hbox{E}\kern-.125emX}}

\newcommand\undermat[2]{
	\makebox[0pt][l]{${\underbrace{\hphantom{
					\begin{matrix}#2\end{matrix}}}_{\text{$#1$}}}$}#2}

\begin{document}

\title{One-Bit Radar Processing with Time-Varying Sampling Thresholds}
\author{
	\IEEEauthorblockN{
		Aria~Ameri$^*$, 
		Arindam~Bose,~\IEEEmembership{Student~Member,~IEEE,}
		Jian~Li,~\IEEEmembership{Fellow,~IEEE,}
		\\
		and Mojtaba~Soltanalian,~\IEEEmembership{Member,~IEEE}
		\thanks{%
			$^*$Corresponding author (e-mail: \textit{aameri2@uic.edu}).
		} 
		\thanks{%
			This work was supported in part by U.S. National Science Foundation Grants CCF-1704401, CCF-1704240, ECCS-1708509, ECCS-1809225, and a 2016 UIC College of Engineering SEED Award. Parts of this work have been presented at IEEE Sensor Array And Multichannel Signal Processing Workshop (SAM), Sheffield, UK, July 2018 \cite{ameri2018one}.
		}
		\thanks{%
			A. Ameri, A. Bose and M. Soltanalian are with the Department of Electrical and Computer Engineering at University of Illinois at Chicago, Chicago, IL 60607 USA (e-mail: aameri2@uic.edu, abose4@uic.edu, msol@uic.edu).
		}
		\thanks{%
			J. Li is with the Department of Electrical and Computer Engineering, University of Florida, Gainesville, FL 32611-6130 USA (e-mail: li@dsp.ufl.edu).
		}
	}
}


\maketitle

\begin{abstract}
	Target parameter estimation in active sensing, and particularly radar signal processing, is a long-standing problem that has been studied extensively.
	In this paper, we propose a novel approach for target parameter estimation in cases where one-bit analog-to-digital-converters (ADCs), also known as signal comparators with time-varying thresholds, are employed to sample the received radar signal instead of high-resolution ADCs.
	The considered problem has potential applications in the design of inexpensive radar and sensing devices in civilian applications, and can likely pave the way for future radar systems employing low-resolution ADCs for faster sampling and high-resolution target determination.
	We formulate the target estimation as a multivariate weighted-least-squares optimization problem that can be solved in a cyclic manner.
	Numerical results are provided to exhibit the effectiveness of the proposed algorithms.
\end{abstract}


\IEEEpeerreviewmaketitle

\section{Introduction}
\label{sec:intro}

	\IEEEPARstart{T}{he} problem of target parameter estimation permeates the field of active sensing and radar signal processing.
	Active sensing systems aim to uncover the location and other useful properties such as velocity information and reflectance properties of a target of interest by dispatching a transmit waveform toward the target and studying the received echo reflected by it.
	For example, the complete dynamics of a moving vehicle including its location and velocity with respect to the observer, can easily be found by simply measuring the difference between the transmitted and received electromagnetic waves in time and frequency domain. 
	Further analysis of the received signal can reveal more information about the target vehicle of interest.

	Since the two world wars, radar systems have been developed, improved, and have made their way into diverse applications such as meteorology \cite{stepanenko1975radar, Radarmeteorology}, air traffic control \cite{482830, nolan2010fundamentals}, structural health monitoring \cite{ihn2008pitch, lynch2004design}, synthetic aperture radar applications \cite{soumekh1999synthetic, curlander1991synthetic}, and underwater sensing \cite{heidemann2012underwater, farr2010integrated}, among others.
	Two major factors in radar signal processing are the design of the transmit signals and receive filters for rejection of clutter and interferences, on which there exists an extensive literature; e.g., see \cite{rummler1967technique, pillai1999optimum, delong1967design, bell1993information, kay2007optimal, stoica2008transmit, stoica2012optimization, soltanalian2013joint, naghsh2014doppler}. 
	The unwanted echoes of the transmit signal that are received as delayed and frequency shifted version of the transmitted signal and are correlated with the main backscattered signal from the target of interest, are generally referred to as clutter.
	Furthermore, noise is the term usually used for signal-independent interference such as effects of adverse jamming signals \cite{he2012waveform} as well as thermal noise and atmospheric disturbances.
	Note that both clutter and noise degrade the accuracy of target parameter estimation; thus, making the receive filter heavily dependent not only on the transmit signal but interference as well.
	A judicious design of both the transmit signal and receive filter in a joint manner can consequently lead to a more accurate estimation of the target parameters in a more tractable computational cost for the radar system.

	One immediate and well-known choice for the receive filter would be the matched filter (MF) that maximizes the signal-to-noise ratio (SNR) in the presence of additive white noise.
	The MF multiplies the received signal with a mirrored and delayed image of the transmitted signal \cite{he2012waveform}. 
	By locating the peak of the output signal, MF discovers the time delay of the received signal, which facilitates the estimation of the distance of the target from the radar, otherwise known as the range.
	On the other hand, a relative difference in motion between the target and the radar results in a Doppler frequency shift in the received signal spectrum.
	In the case of a perceivable Doppler shift in the received signal, a bank of MFs is adopted to estimate the Doppler shift, each of which tuned to a different Doppler frequency \cite{radarSignals}.
	However, MF performs poorly in the presence of interference with arbitrary covariance in the received signal \cite{soltanalian2013joint}.	
	It has been shown in the literature that the effects of the clutter can be mitigated by minimizing the sidelobes of an ambiguity function (AF) \cite{woodward2014probability, 1057703, 4103366}.
	Another line of clutter suppression research can be found in \cite{4749273, 6142119, 8314765}, where the autocorrelation sidelobes of the transmit signal is minimized.
	In addition, the negative effects of interference, especially due to jamming, can be avoided by putting little energy of the transmit signal into the frequency bands where presence of jamming is significant.
	Furthermore, different hardware constraints such as maximum clipping of power amplifiers and analog-to-digital converters (ADC) decrease the performance of MF estimation.
	
	For a more efficient estimation of the target parameters, one can aim to maximize the signal-to-clutter-plus-interference ratio (SCIR) in lieu of SNR.
	Such a scenario arises when the target detection performance of the radar is deteriorated by the clutter or jamming.
	In such cases, one can use a mismatched filter (MMF) instead of an MF \cite{4644058}.
	In comparison with the MF, an MMF allows more degrees of freedom by introducing a receive filter and is not subject to various power constraints of the transmit signal such as constant-modulus or low peak-to-average ratio (PAR).
	Hence, a joint design of the transmit signal and the MMF receive filter can offer a more robust parameter estimation framework \cite{1054205}.

	It is important to note that sampling and quantization of the signal of interest is the first step in digital signal processing.
	The analog to digital conversion ideally requires an infinite number of bits to identically represent the original analog signal, which is not feasible in practice.
	In fact, the aforementioned techniques assume that the received signal is available in full precision.
	The resulting error of quantization can then be modeled as additive noise that usually has little to no impact on algorithms that assume the infinite precision case, provided that the sampling resolution is high enough \cite{gianelli2016one}.
	The signals of interest in many modern applications, albeit, are extremely wide band and may pass through several RF chains that require multitudinous uses of ADCs.
	Such modern applications include spectral sensing for cognitive radio \cite{sun2013wideband, lunden2015spectrum}, cognitive radars \cite{lunden2015spectrum}, radio astronomy \cite{burke2009introduction}, automotive short-range radars \cite{strohm2005development}, driver assistant systems \cite{hasch2012millimeter}, to name a few.
	
	The assumption of high-precision data is, however, not appropriate when the measurements are extremely quantized to very low bit-rates.
	Note that, the cost and the power consumption of ADCs grow exponentially with their number of quantization bits and sampling rate \cite{8645383}.
	Such issues can be mitigated by a reduction in the number of quantization bits. In the most extreme case, the sampling process is carried out by utilizing only \emph{one bit per sample}.
	This can be achieved  by repeatedly comparing the signal of interest with a time-varying threshold (reference) level.
	On the plus side, one-bit comparators can provide extremely high sampling rate and are very cheap and easy to manufacture \cite{8645383}.
	Moreover, the one-bit ADCs operate on very low power and they can significantly reduce the data flow in the system, which further reduces the overall energy consumption. 
	One-bit sampling has been studied from a classical statistical signal processing viewpoint in \cite{ribeiro2006bandwidthi, ribeiro2006bandwidthii, host2000effects, bar2002doa, dabeer2006signal, dabeer2008multivariate, 8683876, 7676417}, a compressive sensing viewpoint in \cite{comp_sense_1, comp_sense_3, comp_sense_4, comp_sense_5,dong2015map}, a sampling and reconstruction viewpoint in \cite{masry1980reconstruction, cvetkovic2000single}. 
	It has been shown in \cite{comp_sense_1, comp_sense_3, comp_sense_4, comp_sense_5} that under certain assumptions, with enough one-bit samples one can recover the full-precision data with bounded error.

	In this paper, we study the radar processing and parameter estimation of both stationary and moving targets using one-bit samplers with time-varying thresholds.
	For both cases of stationary and moving targets, we propose a novel approach that is formulated as minimization of a multivariate weighted-least-squares objective with linear constraints that can be solved in an iterative manner. 
	As stated before, the mentioned approach is cost-effective and is computationally efficient.
	This paper considers an extended problem formulation as compared to \cite{ameri2018one}, in the sense that \cite{ameri2018one} only considers the one-bit radar signal processing for stationary targets while this paper studies the more sophisticated scenario of moving targets in addition to the stationary case, among others.
	To the best of our knowledge, this paper is the first comprehensive work introducing one-bit ADCs and the associated data processing in the context of radar.

	The rest of this paper is organized as follows.
	In Section~\ref{sec:dataModel}, we discuss and formulate the estimation problem in the case of a stationary target. 
	Section~\ref{sec:BussgangApproach} describes a state-of-the-art approach to recover target parameters based on the \textit{Bussgang theorem}.
	The proposed algorithm to estimate the aforementioned parameters is presented in Section~\ref{sec:proposedApproach} for a stationary target.
	In Section~\ref{sec:doppler}, we extend the problem formulation, as well as the estimation algorithm, for parameter estimation in moving target scenarios.
	We further extend the parameter estimation formulations for a stationary target scenario to more advanced setups in Section~\ref{sec:remarks}.
	Numerical results that verify the validity of claims and examine the performance of the proposed algorithms are presented in Section~\ref{sec:numericalResults}.
	Finally, Section~\ref{sec:conclusion} concludes the paper.

	\emph{Notation:} We use bold lowercase letters for vectors and bold uppercase letter for matrices.
	$x_i$ denotes the $i$-th component of the vector $\mathbf{x}$. 
	$(\cdot)^T$ and $(\cdot)^H$ denote the transpose and the conjugate transpose of the vector or matrix argument, respectively. $(\cdot)^*$ denotes the complex conjugate of a complex matrix, vector, or scalar.
	$\|\cdot\|$ denotes the $l_2$ norm of a vector, while $\|\cdot\|_F$ denotes the Frobenius norm of a matrix.
	$\Re(\cdot)$ and $\Im(\cdot)$ are the real and imaginary parts of a complex vector or scalar, respectively.
	Furthermore, the sets of real, complex and natural numbers are denoted by $\realR$, $\complexC$, and $\naturalN$ respectively. 
	$\mathrm{sgn}(\cdot)$ is the element-wise sign operator with an output of $+1$ for non-negative numbers and $-1$ otherwise. 
	Moreover, $\mathbf{Diag}(\cdot)$ and $\calN(\cdot)$ represent the diagonalization and the normalization operator on a matrix argument.
	$\mathbb{E}\{ \cdot \}$ and $\text{Cov}(\cdot)$ denote the expectation and the covariance operator, respectively.
	Finally, The symbol $\odot$ represents the Hadamard product of matrices.

\section{System Model}
\label{sec:dataModel}
	Let
	\begin{align}
		\mathbf{s} =
		\begin{bmatrix}
			s_1 & s_2 & \dots & s_N
		\end{bmatrix} 
		^T \in \complexC^{N}
	\label{eq:s}
	\end{align}
	denote the complex-valued radar transmit sequence of length $N$ that will be used to modulate a train of subpulses \cite{radarSignals}.
	The energy of $\{s_k\}_{k=1}^{N}$ is constrained to be $N$:
	\begin{align}
		\|\mbs\|^2=N
	\end{align}
	without any loss of generality.
	We shall first adopt the discrete data model described in \cite{stoica2012optimization, 1642573} in order to layout the problem formulation for the simpler case of non-moving targets.
	Under the assumptions of negligible intrapulse Doppler shift, and that the sampling is synchronized to the pulse rate, the received discrete-time baseband signal after pulse compression and proper alignment to the range cell of interest, will satisfy the following \cite{4644058, 1642573}:
	\begin{align}\label{eq:y}
		\mby = \alpha_0 \begin{bmatrix}
			s_1 \\ \vdots \\ s_{N-1} \\ s_N 
		\end{bmatrix} + 
		\alpha_1\begin{bmatrix}
			0 \\ s_1 \\ \vdots \\ s_{N-1} 
		\end{bmatrix} + \cdots +
		\alpha_{N-1}\begin{bmatrix}
			0 \\ \vdots \\ 0 \\ s_1
		\end{bmatrix} \nonumber\\
		+\alpha_{-1}\begin{bmatrix}
			s_2 \\\vdots \\ s_N \\ 0 
		\end{bmatrix} + \cdots +
		\alpha_{-N+1} \begin{bmatrix}
			s_N \\ 0 \\\vdots \\ 0 
		\end{bmatrix} +\boldsymbol{\epsilon}
	\end{align}
	where $\alpha_0 \in \mathbb{C}$ is the scattering coefficient of the current range cell, $\{\alpha_k\}_{k \neq 0}$ are the scattering coefficients of the adjacent range cells that contribute to the clutter in $\mathbf{y}$, and $\boldsymbol{\epsilon}$ is the signal-independent noise which comprises of measurement noise as well as other disturbances such as jamming.
	By collecting all the delayed samples of the transmitted signal into a matrix, the data model in \eqref{eq:y} can be simplified as
	\begin{align}
		\mathbf{y}=\mathbf{A}^H\boldsymbol{\alpha}+\boldsymbol{\epsilon}
	\label{eq:y2}
	\end{align}
	where
	\begin{align}
		\mathbf{A}^H&=
		\begin{bmatrix}
			s_1 	  	& 0 		 	& \dots 	& 0 		 	& s_N 	  	& s_{N-1} 	& \dots 	& s_2 		\\
			s_2 	  	& s_1 			& \dots 	& \vdots 		& 0 	 	& s_N 		& \dots 	& \vdots 	\\
			\vdots 		& \vdots	 	& \ddots	& 0 		  	& \vdots  	& \vdots 	& \ddots 	& s_N		\\
			s_N 	  	& s_{N-1}  		& \dots 	& s_1 		   	& 0 		& 0 		& \dots 	& 0 		\\
		\end{bmatrix}, 
		\label{eq:A} 
	\end{align} and
	\begin{align}
		\boldsymbol{\alpha} =
		\left[\alpha_0, \alpha_1, \dots, \alpha_{N-1}, \alpha_{-(N-1)}, \dots, \alpha_{-1}
		\right]^T
		\label{eq:alpha}
	\end{align} is the corresponding scattering coefficient vector. 
	In \eqref{eq:A}, the first column of $\mbA^H$ represents the principal reflection from the target after range cell alignment, and the second to the last column of the same are in fact the different delayed echos of the transmit signal $\mbs$ (see \cite{1642573} for more details).
	Furthermore, if the Doppler shifts are not negligible due to the relative difference in motion between the target and the radar system, the data model in \eqref{eq:y2} needs to be modified to accommodate the same, and has been discussed in Section~\ref{sec:doppler}.

	By applying one-bit comparators at the receiver, the sampled baseband signal can be written as:
	\begin{align}
		\begin{tabular}{c}
			$\boldsymbol{\gamma}_r=\mathrm{sgn}\left(\Re\{\mathbf{A}^H\boldsymbol{\alpha}+\boldsymbol{\epsilon}-\boldsymbol{\lambda}\}\right)$, \\
			$\boldsymbol{\gamma}_i=\mathrm{sgn}\left(\Im\{\mathbf{A}^H\boldsymbol{\alpha}+\boldsymbol{\epsilon}-\boldsymbol{\lambda}\}\right)$,\\
			$\boldsymbol{\gamma}=\frac{1}{\sqrt{2}} (\boldsymbol{\gamma}_r + j \boldsymbol{\gamma}_i)$,
		\end{tabular}
		\label{eq:gamma}
	\end{align}
	where $\boldsymbol{\lambda}$ is the tunable complex-valued threshold level vector at the comparators, whose design is discussed in Section~\ref{sec:proposedApproach}.
	Note that in \eqref{eq:gamma}, we sample both real and imaginary parts of the received signal in order to preserve the phase information.
	We further assume that the clutter  coefficients $\{\alpha_k\}_{k \neq 0}$ are zero-mean and their variance,
	\begin{align}
	\beta \triangleq \mathbb{E}\{\left|\alpha_k\right|^2\}, \qquad k \neq 0,
	\end{align}
	and the covariance matrix of $\boldsymbol{\epsilon}$,
	\begin{align}
		\bGamma \triangleq \mathbb{E}\{\bepsilon\bepsilon^H\},
	\label{eq:Gamma}	
	\end{align}
	are known quantities.
	We further assume that $\{\alpha_k\}_{k \neq 0}$ are independent of each other and of $\boldsymbol{\epsilon}$ as well.
	Note that, in radar applications, both $\boldsymbol{\Gamma}$ and $\beta$ can be acquired using various preprocessing techniques, e.g. by employing pre-scans, and are typically assumed to be known \textit{a priori} \cite{stoica2012optimization}. A detailed discussion of the prescanning process can be found in \cite{cognitiveRadar}.

	As mentioned earlier, once the received signal $\mathbf{y}$ is available, one can  estimate the target backscattering coefficient $\alpha_0$ by exploiting the signal model in \eqref{eq:y2} using a mismatched filter (MMF).
	The MMF estimate of $\alpha_0$ has the following linear form in $\mathbf{y}$ \cite{soltanalian2013joint}:
	\begin{align}
		\hat{\alpha}_0=\frac{\mathbf{w}^H\mathbf{y}}{\mathbf{w}^H\mathbf{s}}
	\label{eq:alpha_0_estimate}
	\end{align}
	where $\mathbf{w} \in \mathbb{C}^N$ is the MMF vector of the receive filter.
	The mean squared error (MSE) of the mentioned estimate can be derived as
	\begin{align}
		\text{MSE}(\hat{\alpha}_0)=\mathbb{E}\left\{\left|\frac{\mathbf{w}^H\mathbf{y}}{\mathbf{w}^H\mathbf{s}} - \alpha_0\right|^2\right\}=\frac{\mathbf{w}^H\mathbf{R}\mathbf{w}}{\left|\mathbf{w}^H\mathbf{s}\right|^2}
	\label{eq:MSE}
	\end{align}
	where
	\begin{align}
		\mathbf{R}=\beta \sum_{0< |k|  \leq (N-1)} \mathbf{J}_k\mathbf{s}\mathbf{s}^H\mathbf{J}_k^H + \boldsymbol{\Gamma},
	\end{align}
	is the covariance matrix of the interference terms in \eqref{eq:y2} and $\{\mathbf{J}_k\}$ are the shift matrices formulated as
	\begin{align}
		\mathbf{J}_k=\mathbf{J}_{-k}^H=
		\begin{bmatrix}
			0 			& \dots & 0 	& 1 	& \dots 	& 0   	\\
			\vdots		& 		&	 	&	  	& \ddots 	& 		\\
			& 		&    	&	 	&			& 1  	\\
			\undermat{k}{0 & \dots 	& 0} 	& \dots	 	&	& 	 	\\
			\end{bmatrix}^H_{N \times N}, \\
			k = 0, 1, \cdots, N-1. \nonumber
		\label{eq:J_k}	
	\end{align}
	Note that the denominator of the MSE in \eqref{eq:MSE} is the power of the signal at the receiver and its numerator is the power of the interferences.
	Therefore, minimizing the MSE is identical to maximizing the SCIR.

	Note that one can exploit the relationship between the covariance matrices of the received signals before and after the non-linear transformation of one-bit sampling in order to estimate the target parameter $\alpha_0$.
	This relationship is provided by the Bussgang theorem in a \textit{normalized} sense \cite{bussgang1952crosscorrelation}.
	In the following section, we briefly discuss a state-of-the-art Bussgang-theorem-aided procedure to estimate $\alpha_0$. 
	Afterwards, in Section~\ref{sec:proposedApproach}, we propose an algorithm that, through minimizing the MSE, jointly recovers the scattering coefficient of the current range cell, $\alpha_0$, and the received signal, $\mathbf{y}$, from the one-bit sampled received data $\boldsymbol{\gamma}$, as introduced in \eqref{eq:gamma}.

\section{Bussgang-Theorem-Aided Estimation}
\label{sec:BussgangApproach}
	In this section, we describe a state-of-the-art Bussgang-theorem-aided approach to estimate the target parameters \cite{bussgang1952crosscorrelation}.
	Let $Y(t)$ be a real-valued, scalar, and stationary Gaussian process that undergoes the one-bit sampling process $Z(t)=\text{sgn}(Y(t))$.
	The autocorrelation function of the process $Z(t)$, denoted by $R_Z(\tau)$, is given by
	\begin{align}
		R_Z(\tau) = \mathbb{E} \{ Z(t+\tau) Z(t) \} = \frac{2}{\pi} \asin \left( \bar{R}_Y(\tau) \right)
	\end{align}
	where $\bar{R}_Y(\tau) = R_Y(\tau) / R_Y(0)$ is the normalized autocorrelation function of the process $Y(t)$ \cite{van1966spectrum}.
	The good news is that the Bussgang theorem \cite{bussgang1952crosscorrelation} states that the crosscorrelation function of the processes $Y(t)$ and $Z(t)$ is proportional to the autocorrelation function of $Y(t)$, i.e. $R_{ZY}(\tau) = \mu R_Y(\tau)$, where the factor $\mu$ depends on the power of the process $Y(t)$. 

	The case of complex-valued vector processes, which was studied in \cite{liu2017one}, can be extended in a similar manner.
	Let $\mathbf{y}$ be the complex-valued vector whose one-bit samples are given by $\boldsymbol{\gamma} = \frac{1}{\sqrt{2}} \left( \mathrm{sgn}(\Re(\mathbf{y})) + j \mathrm{sgn}(\Im(\mathbf{y})) \right)$, as in~\eqref{eq:gamma}.
	Then the normalized autocorrelation of the vector $\mathbf{y}$ is given by
	\begin{align}
		\bar{\mathbf{R}}_\mathbf{y} = \mathcal{N}(\mathbf{R}_\mathbf{y}) \triangleq \mathbf{D}^{-1/2} \mathbf{R}_\mathbf{y} \mathbf{D}^{-1/2}
	\end{align}
	where $\mathbf{D} = \mathbf{R}_\mathbf{y} \odot \mathbf{I}$ is a diagonal matrix containing only the diagonal entries of $\mathbf{R}_\mathbf{y}$.
	It has been shown in \cite{liu2017one} that the following covariance equality holds:
	\begin{align}
		\bar{\mathbf{R}}_\mathbf{y}  =  \sin \left( \frac{\pi}{2} \mathbf{R}_{\boldsymbol{\gamma}} \right),
		\label{eq:RBussgang}
	\end{align}
	where $\mathbf{R}_{\boldsymbol{\gamma}}$ is the autocorrelation matrix of the one-bit sampled data, $\bgamma$.

	In order to apply the above results to the one-bit radar processing problem using a threshold level vector $\blambda \ \in \complexC^N$, we can derive the covariance matrix of the difference between the received signal and the time-varying threshold, viz.
	\begin{align}
		\mathbf{R}_{\mathbf{y} - \boldsymbol{\lambda}} = |\alpha_0|^2 \mathbf{s}\mathbf{s}^H + \boldsymbol{\lambda}\boldsymbol{\lambda}^H + \mathbf{R} - 2\Re(\alpha_0 \mathbf{s} \boldsymbol{\lambda}^H).
	\end{align}
	Therefore, one can compute the scattering coefficient $\alpha_0$, by solving the following non-convex optimization problem:
	\begin{align}
		\underset{\alpha_0}{\text{min}} \quad
		\left\| \bar{\mathbf{R}}_{\mathbf{y} - \boldsymbol{\lambda}} - \mathcal{N} ( |\alpha_0|^2 \mathbf{s}\mathbf{s}^H + \boldsymbol{\lambda}\boldsymbol{\lambda}^H + \mathbf{R} - 2\Re(\alpha_0 \mathbf{s} \boldsymbol{\lambda}^H) ) \right\|_F
	\end{align}
	in which $\bar{\mathbf{R}}_{\mathbf{y} - \boldsymbol{\lambda}}$ is obtained via \eqref{eq:RBussgang}, and using only one observation or \textit{snapshot} of $\boldsymbol{\gamma}$.
	
\section{The Proposed Approach for Stationary Targets}
\label{sec:proposedApproach}
	In this section, we address the proposed approach to recover both the received signal $\mathbf{y}$ and the scattering coefficient $\alpha_0$ from the one-bit sampled received signal $\boldsymbol{\gamma}$ for a stationary target by minimizing the aforementioned MSE in \eqref{eq:MSE}. 
	
	For a given transmit sequence $\mathbf{s}$, the optimum receive filter $\mathbf{w}$ can be simply given as closed form solution \cite{soltanalian2013joint,stoica2012optimization}:
	\begin{align}
		\mathbf{w}=\mathbf{R}^{-1}\mathbf{s}
		\label{eq:optimal_w_MMF}
	\end{align}
	up to a multiplicative constant.
	Nevertheless, the MMF approach to recover $\alpha_0$, discussed in \eqref{eq:alpha_0_estimate}, requires the availability of the un-quantized (or high-resolution quantized) received signal $\mathbf{y}$, which is unfortunately not available directly due to the one-bit sampling of the received signal.
	Therefore, we shall resort to an alternative optimization approach that utilizes the one-bit sampled data $\boldsymbol{\gamma}$ in lieu of $\mathbf{y}$ in order to estimate the target parameter.
	In pursuance of radar parameter recovery using one-bit sampled data with time-varying thresholds, we analyze two matters of major significance:  
	\begin{enumerate*}[label=(\roman*)]
		\item the recovery of $\mathbf{y}$ and estimation of $\alpha_0$ by employing the one-bit data procured at the receiver,
		and 
		\item the design of next set of thresholds to be used at the one-bit ADCs.
	\end{enumerate*}

	\subsection{Estimation of Target Parameters}
	\label{subsec:estimate}
		In order to efficiently estimate the received signal $ \mathbf{y} $ and target parameter $ \alpha_0 $, we consider minimizing of the following weighted-least-squares (WLS) objective in a more generalized sense:
		\begin{align}
			Q(\mathbf{y}, \alpha_0) \triangleq (\mathbf{y}-\alpha_0\mathbf{s})^H \mathbf{R}^{-1} (\mathbf{y}-\alpha_0\mathbf{s}).
			\label{eq:Q}
		\end{align}
		It should be noted that the usage of the above criterion has following advantages: 
		\begin{enumerate}
			\item Unlike the MMF in \eqref{eq:alpha_0_estimate}, $Q$ does not require a knowledge of $\mathbf{y}$.
			\item It is a function of both $\mathbf{y}$ and $\alpha_0$, laying the ground for their respective joint recovery.
			\item It can easily be observed that, for any given $\mathbf{y}$, the optimum $\alpha_0$ in \eqref{eq:Q} is identical to that of MMF in \eqref{eq:alpha_0_estimate} with the use of \eqref{eq:optimal_w_MMF}--- thus making it a natural choice for parameter recovery.
			\label{enum:alpha_0_optimal}
			\item In effect, the minimization of \eqref{eq:Q} enforces the system model introduced in \eqref{eq:y}.
			Note that the model mismatch can be written as 
			\begin{align}
				\mathbf{y} - \alpha_0 \mathbf{s} = \tilde{\mathbf{A}}^H\tilde{\boldsymbol{\alpha}} + \boldsymbol{\epsilon}
			\end{align}
			where $\tilde{\mathbf{A}}^H$ and $\tilde{\boldsymbol{\alpha}}$ are derived from $\mathbf{A}^H$ and $\boldsymbol{\alpha}$ with their first column and first entry dropped, respectively.
			It can easily be established that the objective function in \eqref{eq:Q} penalizes the model mismatch based on the second order mismatch statistics derived as
			\begin{eqnarray}
				& &\mathbb{E} \left\{ \left( \tilde{\mathbf{A}}^H\tilde{\boldsymbol{\alpha}} + \boldsymbol{\epsilon} \right) \left( \tilde{\mathbf{A}}^H\tilde{\boldsymbol{\alpha}} + \boldsymbol{\epsilon} \right)^H \right\} \nonumber \\
				&= &\mathbb{E} \left\{ \tilde{\mathbf{A}}^H\tilde{\boldsymbol{\alpha}}\tilde{\boldsymbol{\alpha}}^H\tilde{\mathbf{A}}  \right\} + \mathbb{E} \left\{ \boldsymbol{\epsilon}\boldsymbol{\epsilon}^H  \right\} \nonumber \\
				&= &\mathbf{R}.
			\end{eqnarray}
		\end{enumerate}
	
		Hence, based on the property 3, and by substituting the optimum $\alpha_0$, the objective function \eqref{eq:Q} can be reformulated as
		\begin{align}
			Q(\mathbf{y}, \hat{\alpha}_0) &\triangleq Q(\mathbf{y}) \nonumber \\
			&= \mathbf{y}^H \left( \mathbf{I} - \frac{\mathbf{s}\mathbf{w}^H}{\mathbf{w}^H\mathbf{s}} \right)^H \mathbf{R}^{-1} \left( \mathbf{I} - \frac{\mathbf{s}\mathbf{w}^H}{\mathbf{w}^H\mathbf{s}} \right) \mathbf{y}. 
		\end{align}
		Hence, the problem of jointly estimating $\alpha_0$ and $\mathbf{y}$ boils down to:
		\begin{align}
			\label{eq:opt_problem}
			& \underset{\mathbf{y}}{\text{min}} & & \mathbf{y}^H \left[ \left( \mathbf{I} - \frac{\mathbf{s}\mathbf{w}^H}{\mathbf{w}^H\mathbf{s}} \right)^H \mathbf{R}^{-1} \left( \mathbf{I} - \frac{\mathbf{s}\mathbf{w}^H}{\mathbf{w}^H\mathbf{s}} \right) \right] \mathbf{y} \\
			& \text{s.t.} & & 
			\boldsymbol{\Omega}_r \left( \mathbf{y}_r - \boldsymbol{\lambda}_r \right) \geq \mathbf{0}, \nonumber \\
			& & & \boldsymbol{\Omega}_i \left( \mathbf{y}_i - \boldsymbol{\lambda}_i \right) \geq \mathbf{0}, \nonumber
		\end{align}
		where $(\mathbf{y}_r, \mathbf{y}_i)$ and $(\boldsymbol{\lambda}_r,\boldsymbol{\lambda}_i)$ denote the real and imaginary parts of $\mathbf{y}$ and  $\boldsymbol{\lambda}$, respectively, and $\boldsymbol{\Omega}_r \triangleq \mathbf{Diag}(\boldsymbol{\gamma}_r)$,  $\boldsymbol{\Omega}_i  \triangleq \mathbf{Diag}(\boldsymbol{\gamma}_i)$.
		One can easily verify that the optimization problem in \eqref{eq:opt_problem} is a convex quadratic program with linear constraints  that can be solved efficiently.
		Upon finding the optimal $\mathbf{y}$, the optimal $\alpha_0$ can be calculated using the MMF estimate in \eqref{eq:alpha_0_estimate}.
	
		\begin{algorithm}[t]
			\caption{One-Bit Radar Parameter Estimation for Stationary Targets}\label{algorithm:opt}
			\begin{algorithmic}[1]
				\Ensure The transmit sequence $\mathbf{s}$ and set the threshold vector $\boldsymbol{\lambda}$ arbitrarily, or generate according to \eqref{eq:lambda_set}-\eqref{eq:lambda_char}.
				
				\State Compute the optimal MMF vector $\mathbf{w}$ using \eqref{eq:optimal_w_MMF}.
				
				\State Compute the optimal vector $\mathbf{y}$ by solving \eqref{eq:opt_problem}.
				
				\State Estimate the target scattering coefficient $\alpha_0$ using the MMF estimator in \eqref{eq:alpha_0_estimate}.
				
				\State In case of tracking, set $\boldsymbol{\lambda}$ according to \eqref{eq:lambda_set}-\eqref{eq:lambda_char} and go to Step 2.
			\end{algorithmic}
		\end{algorithm}

	\subsection{Time-Varying Threshold Design}
	\label{subsec:threshold}
		\subsubsection{Sampling with a Single One-Bit ADC} 
			From an information theoretic viewpoint, in order to collect the \emph{most} information on $\mathbf{y}$, one could expect $\boldsymbol{\lambda}$ to be set in such a way that by considering the \emph{a priori} information, observing any of the two outcomes in the set $\{-1,+1\}$ at the output of the one-bit sampler for a single sample has the same likelihood.
			In a general case, $\boldsymbol{\lambda}$ is expected to partition the set of likely events into subsets with similar \emph{cardinality}.
			When the probability density function (pdf) of the received signal follows a Gaussian distribution, this goal is achieved by setting $\boldsymbol{\lambda}$  as close as possible to the expected value of $\mathbf{y}$.
			More precisely, we choose:
			\begin{align}
				\boldsymbol{\lambda}= \mathbb{E}\left\{\alpha_0\right\} \mathbf{s} .
				\label{eq:lambda_set}
			\end{align}
			In other words, the choice of $\boldsymbol{\lambda}$  will be governed by our \textit{future} expectation of the value of $\alpha_0$.
			This is particularly pertinent to target tracking scenarios.  

\vspace{5pt}
		\subsubsection{Sampling with Multiple One-Bit ADCs} 
			We note that our estimation method can easily be extended to cases where the signal is sampled by several ADCs in parallel.
			This only leads to extra linear constraints in \eqref{eq:opt_problem}.
			Assuming that $K$ number of ADCs are used and the thresholds are set \emph{a priori}, in the single sample case, the thresholds are optimal if they partition the set of likely events into $K+1$ subsets with similar \emph{cardinality}.
			The determination of the thresholds will be even more difficult when the number of samples or the number of ADCs grow large. 
			However, a close \emph{approximation} of the optimal threshold vectors $\{\boldsymbol{\lambda_k}\}_{k=1}^K$ can be obtained by assuming $\{\boldsymbol{\lambda_k}\}_{k=1}^K$ to be random variables \cite{8645383}.
			In other words, a good set of random sampling threshold vectors $\{\boldsymbol{\lambda_k}\}_{k=1}^K$ should mimic the behavior of~$\mathbf{y}$.
			In particular, we generate $\{\boldsymbol{\lambda_k}\}_{k=1}^K$  as a set of random vectors similar to $\mathbf{y}$ that has the same (Gaussian) distribution:
			\begin{eqnarray}
				\mathbb{E}\left\{\boldsymbol{\lambda}\right\} &=& \mathbb{E}\left\{\alpha_0\right\} \mathbf{s}, \nonumber \\
				\text{Cov}(\boldsymbol{\lambda}) &=& \mathbb{E}\left\{|\alpha_0|^2 \right\} \, \mathbf{s}  \mathbf{s} ^H+\mathbf{R}.
				\label{eq:lambda_char}
			\end{eqnarray}
	
			The steps of the proposed approach are summarized in Algorithm~\ref{algorithm:opt} for reader’s convenience.

\section{Parameter Estimation for Moving Targets}
\label{sec:doppler}
	In this section, we consider the moving targets scenario where the Doppler effect can no longer be neglected.
	In order to perform a recovery of radar parameters, i.e. the backscattering coefficient and the Doppler shift of the target, we first update the system model of \eqref{eq:y} and then modify the proposed approach discussed in Section~\ref{sec:proposedApproach} to recover the normalized Doppler shift as well. 

	\subsection{Modified Problem Formulation}
	\label{subsec:movingTargetModel}
\begin{figure}
		\centering
		\includegraphics[width=.7\linewidth,draft=false]{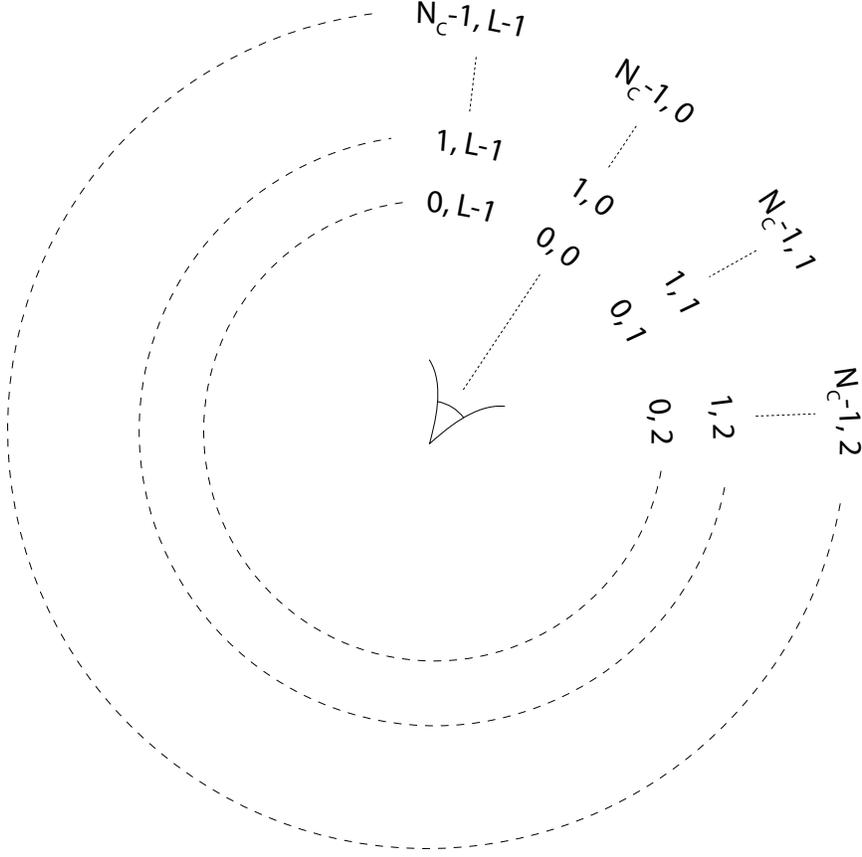}
		\caption{The setting for different range-azimuth cells. All the cell numbers are shown in (range, azimuth) pairs.}
	\label{fig:rangeAzimuth}	
	\end{figure}

		Let $\mathbf{s} \in \mathbb{C}^N$ denote the discrete-time transmit sequence of a digital system, as in \eqref{eq:s}. 
		After alignment to the range-azimuth cell of interest, the new discrete-time complex-valued received baseband data vector, which is backscattered from the moving target in the corresponding range-azimuth cell, can be formulated as (see \cite{naghsh2014doppler, 6407265, gini2012waveform, 6404093})
	\begin{align}
		\mathbf{y} = \alpha_0 (\mathbf{s} \odot \mathbf{p}(\nu)) + \mathbf{c} + \mathbf{n},
		\label{eq:modelDoppler}
	\end{align}
	where $\alpha_0 $ is the complex backscattering coefficient of the target in the current range-azimuth cell and $\mathbf{p}(\nu) = \begin{bmatrix}
	e^{j2\pi(0)\nu}, & e^{j2\pi(1)\nu}, & \dots, & e^{j2\pi(N-1)\nu}
	\end{bmatrix}^T$ is the propagation effect vector with $\nu \in \left[-.5, 5\right)$ being the normalized Doppler shift of the target.
	The $N$-dimentional vectors $\mathbf{c}$ and $\mathbf{n}$ denote the signal-dependent clutter and signal-independent noise, respectively.
	The clutter vector $\mathbf{c}$ is comprised of returned echos from uncorrelated scatterers at different range-azimuth cells \cite{6404093} (as depicted in Fig.~\ref{fig:rangeAzimuth}), which are spread in Doppler frequency due to the possible clutter motion and can be formulated as 
	\begin{align}
		\mathbf{c} = \sum_{k=0}^{N_c-1} \sum_{l=0}^{L-1} \alpha_{(k,l)} \mathbf{J}_k \left[ \mathbf{s} \odot \mathbf{p}(\nu_{d_{(k,l)}}) \right],
	\end{align}
	where $N_c \leq N$ is the number of range-rings, $L$ is the number of various azimuth sectors, and $\alpha_{(k,l)}$ and $\nu_{d_{(k,l)}}$ are the scattering coefficient and normalized Doppler shift of the $(k,l)$-th range-azimuth cell, interfering with the range-azimuth cell of interest.
	The matrix $\mathbf{J}_k$ is defined in the same way as in \eqref{eq:J_k}.

	The covariance matrix of the clutter vector $\mathbf{c}$ can be written as
	\begin{align} \label{eq:Sigmac}
		\boldsymbol{\Sigma}_{\mathbf{c}} = \sum_{k=0}^{N_c-1} \sum_{l=0}^{L-1} \sigma_{(k,l)}^2 \mathbf{J}_k \boldsymbol{\Phi}\left( s, (k,l) \right) \mathbf{J}_k^T
	\end{align}
	with $\sigma_{(k,l)}^2$ being the average scattering power of the scatterer in $(k,l)$-th range-azimuth cell.
	The clutter patches in each range-azimuth cell are assumed to have uniform Doppler shifts in the interval $\Omega_c = \left( \bar{\nu}_{d_{(k,l)}} - \frac{\epsilon_{d_{(k,l)}}}{2}, \bar{\nu}_{d_{(k,l)}} + \frac{\epsilon_{d_{(k,l)}}}{2} \right)$~\cite{gini2012waveform}. 
	Note that the assumption of having uniform Doppler shifts in each range-azimuth cell, results from the fact that the clutter patches in these cells can be any object in our environment, with some of them moving. 
	If such objects are moving, even slightly, the echoes reflected from the corresponding cells will have a Doppler shift associated with that movement.
	Examples of such objects include vehicles, ocean waves, and trees with moving leaves due to the wind \cite{riverSurfaceCurrents, 923012}.
	These contribute to a small Doppler frequency shift in clutter input which is assumed to be distributed uniformly. 
	
	Moreover, $\boldsymbol{\Phi}$ in \eqref{eq:Sigmac} can be expressed as
	\begin{align*}
		\boldsymbol{\Phi}\left( s, (k,l) \right) = \mathbf{Diag}(\mathbf{s}) \mathbf{C}_\nu(k, l) \mathbf{Diag}(\mathbf{s})^H,
	\end{align*}
	where $\mathbf{C}_\nu(k, l)$ is the covariance matrix of the propagation effect vector of the $(k,l)$-th cell \cite{6404093}, defined as
	\begin{align}
		\mathbf{C}_\nu(k, l) = 
		\begin{cases}
			1														
					& k = l\\
			e^{j(k-l)\bar{\nu}_{d_{(k,l)}}}	\frac{\sin\left(\frac{k-l}{2}\epsilon_{d_{(k,l)}}\right)}{\sin\left(\frac{k-l}{2}\epsilon_{d_{(k,l)}}\right)}						
					& k \neq l
		\end{cases}.
	\end{align}
	Similar to \eqref{eq:Gamma}, we denote the covariance matrix of the signal-independent interference by $\boldsymbol{\Gamma}$, and redefine the covariance matrix of the interference as follows:
	\begin{align}
		\mathbf{R} = \text{Cov} (\mathbf{c} + \mathbf{n}) = \boldsymbol{\Sigma}_{\mathbf{c}} + \boldsymbol{\Gamma}.
		\label{eq:RDoppler}
	\end{align}

	\begin{algorithm}[t]
		\caption{One-Bit Radar Parameter Estimation for Moving Targets}\label{algorithm:optDoppler}
		\begin{algorithmic}[1]
			\Ensure The transmit sequence $\mathbf{s}$ and set the threshold vector $\boldsymbol{\lambda}$ arbitrarily or generate according to \eqref{eq:lambda_char_Doppler}.
			
			\State For fixed $\mathbf{y}$, $\alpha_0$, and $\nu$, compute the optimal MMF vector $\mathbf{w}$ according to \eqref{eq:optimal_w_MMF_Doppler}.
			
			\State For fixed $\mathbf{w}$ and $\nu$, compute the optimal vector $\mathbf{y}$ by solving the criterion in  \eqref{eq:opt_problem_doppler} with respect to $\mathbf{y}$.
			
			\State For fixed $\mathbf{y}$ and $\mathbf{w}$, compute the optimal target normalized Doppler shift $\nu$ by minimizing the criterion in \eqref{eq:opt_problem_doppler_nu}.
			
			\State If convergence is reached, go to Step 5; otherwise, go to Step 1.
			
			\State For fixed $\mathbf{w}$, $\mathbf{y}$, and $\nu$, estimate the target backscattering coefficient $\alpha_0$ using \eqref{eq:alpha_0_estimate_doppler}.
			
			\State In case of tracking, set $\boldsymbol{\lambda}$ according to \eqref{eq:lambda_char_Doppler} and go to Step 1. \looseness=-1
		\end{algorithmic}
	\end{algorithm}

	\subsection{Estimation of Target Parameters}
	\label{subsec:movingTargetEstimation}
		When the received signal is available and the Doppler shift is known, an estimation of the backscattering coefficient $\alpha_0$ with minimal MSE can be achieved by using a mismatched filter, in a similar manner as in \eqref{eq:alpha_0_estimate}.
		The estimate of the target backscattering coefficient given by MMF is
		\begin{align}
			\hat{\alpha}_0=\frac{\mathbf{w}^H\mathbf{y}}{\mathbf{w}^H \left(\mathbf{s} \odot \mathbf{p}(\nu)\right)}.
			\label{eq:alpha_0_estimate_doppler}
		\end{align}
		Additionally, it can be verified that the optimal $\mathbf{w}$ that minimizes the MSE criterion is given by
		\begin{align}
			\mathbf{w}=\mathbf{R}^{-1}\left(\mathbf{s} \odot \mathbf{p}(\nu)\right).
			\label{eq:optimal_w_MMF_Doppler}	
		\end{align}
		up to a multiplicative constant.

		Note once again that, due to using one-bit ADCs, the access to the received signal $\mathbf{y}$ is restricted to only its one-bit samples, given by \eqref{eq:gamma}.
		In order to tackle the problem of estimating the backscattering coefficient $\alpha_0$ and the normalized Doppler shift $\nu$, we form a modified version of the weighted-least-squares objective function in \eqref{eq:Q}, in compliance with the system model defined in \eqref{eq:modelDoppler}:
		\begin{align}
			\label{eq:QDoppler}	
			\tilde{Q}(\mathbf{y}, &\alpha_0, \nu) \triangleq \\
			& \left[\mathbf{y}- \alpha_0 (\mathbf{s} \odot \mathbf{p}(\nu)) \right]^H \mathbf{R}^{-1} \left[\mathbf{y}- \alpha_0 (\mathbf{s} \odot \mathbf{p}(\nu)) \right] \nonumber
		\end{align}
		with $\mathbf{R}$ being the covariance matrix of the interference defined in \eqref{eq:RDoppler}.
		
		Similar to the stationary target case, the aforementioned objective function is chosen to have the following properties: \begin{enumerate}
			\item It does not rely on the knowledge of the received signal $\mathbf{y}$ and yet enforces the system model in \eqref{eq:modelDoppler}, 
			\item For given $\mathbf{y}$ and $\nu$, the optimal $\alpha_0$ of \eqref{eq:QDoppler} is identical to the MMF estimate of $\alpha_0$ in \eqref{eq:alpha_0_estimate_doppler}, 
			\item It is a function of $\mathbf{y}$, $\alpha_0$, and $\nu$, which permits their joint estimation,
			and last but not least,
			\item The recovery of $\mathbf{y}$ using $\tilde{Q}$ paves the way for usage of other classical signal processing methods that rely on the knowledge of $\mathbf{y}$.
		\end{enumerate}
	
		The problem of jointly estimating $\mathbf{y}$, $\alpha_0$, and $\nu$ for moving target determination thus becomes 
		\begin{align}
			& \underset{\mathbf{y}, \alpha_0, \nu}{\text{min}} & & \left[\mathbf{y}- \alpha_0 (\mathbf{s} \odot \mathbf{p}(\nu)) \right]^H \mathbf{R}^{-1} \left[\mathbf{y}- \alpha_0 (\mathbf{s} \odot \mathbf{p}(\nu)) \right] \nonumber \\
			& \text{s.t.} & & 
			\boldsymbol{\Omega}_r \left( \mathbf{y}_r - \boldsymbol{\lambda}_r \right) \geq \mathbf{0}, \nonumber \\
			& & & \boldsymbol{\Omega}_i \left( \mathbf{y}_i - \boldsymbol{\lambda}_i \right) \geq \mathbf{0}. 
			\label{eq:opt_problem_doppler}	
		\end{align}
		However, by substituting the optimal $\alpha_0$ in \eqref{eq:alpha_0_estimate_doppler} into the objective function of \eqref{eq:opt_problem_doppler}, we achieve a more simplified optimization problem:
		\begin{align}
			& \underset{\mathbf{w}, \mathbf{y}, \nu}{\text{min}} & & \left\| \mathbf{R}^{-1/2} \left( \mathbf{I} - \frac{[\mathbf{s} \odot \mathbf{p}(\nu)]\mathbf{w}^H}{\mathbf{w}^H[\mathbf{s} \odot \mathbf{p}(\nu)]} \right) \mathbf{y} \right\|^2_2 \nonumber \\
			& \text{s.t.} & & 
			\boldsymbol{\Omega}_r \left( \mathbf{y}_r - \boldsymbol{\lambda}_r \right) \geq \mathbf{0}, \nonumber \\
			& & & \boldsymbol{\Omega}_i \left( \mathbf{y}_i - \boldsymbol{\lambda}_i \right) \geq \mathbf{0}. 
			\label{eq:optProblemY}	
		\end{align}
		In order to solve the above minimization problem, we resort to cyclic optimization over $\mathbf{w}$, $\mathbf{y}$, and $\nu$, until convergence.
		The optimal $\mathbf{w}$ for fixed $\mathbf{y}$ and $\nu$ can be obtained using \eqref{eq:optimal_w_MMF_Doppler}.
		Next, for fixed $\mathbf{w}$ and $\nu$, it is easy to see that the above optimization problem is a convex linearly-constrained quadratic program with respect to $\mathbf{y}$, which can be efficiently solved.
		Lastly, in order to find the optimal normalized Doppler shift $\nu$ when $\mathbf{w}$ and $\mathbf{y}$ are fixed, we can rewrite the optimization problem \eqref{eq:opt_problem_doppler} with respect to $\nu$ as:
		\begin{align}
			\label{eq:opt_problem_doppler_nu}
			& \underset{\nu}{\text{min}} & & g(\nu) \\
			& \text{s.t.} & & 
			\mathbf{p}(\nu) = \begin{bmatrix}
			e^{j2\pi (0)\nu} & e^{j2\pi (1)\nu} & \dots & e^{j2\pi (N-1)\nu} 
			\end{bmatrix}^T \nonumber
		\end{align}
		where \\
		\resizebox{\linewidth}{!}{
		\begin{minipage}{1\linewidth}
			\begin{align*}
				& g(\nu) \triangleq \\
				& \begin{bmatrix}
				1 \\ \mathbf{p}(\nu)
				\end{bmatrix}^H
				\begin{bmatrix}
				0 & -(\hat{\alpha}_0 \mathbf{s})^T \odot (\mathbf{y}^H  \mathbf{R}^{-1}) \\
				-(\hat{\alpha}_0 \mathbf{s})^*  \odot (\mathbf{R}^{-1} \mathbf{y}) & 
				|\hat{\alpha}_0|^2 \mathbf{R}^{-1} \odot (\mathbf{s} \mathbf{s}^H)^*
				\end{bmatrix}
				\begin{bmatrix}
				1 \\ \mathbf{p}(\nu)
				\end{bmatrix}
			\end{align*}
		\end{minipage}
		} \vspace{.5em} \\
		and where $\hat{\alpha}_0$ is calculated using \eqref{eq:alpha_0_estimate_doppler}.

		The optimization problem in \eqref{eq:opt_problem_doppler_nu} resembles that of estimating the direction-of-arrival (DOA) in uniform linear arrays (ULAs) and can be dealt with using one of the many algorithms for estimating the DOA--- see \cite{stoica2005spectral} for details.
		We repeat the cyclic optimization procedure until a pre-defined convergence criterion is satisfied.
		Once the $\mathbf{w}$, $\mathbf{y}$, and $\nu$ are estimated, the backscattering coefficient $\alpha_0$ can be easily retrieved via \eqref{eq:alpha_0_estimate_doppler}.

		As to the design of the threshold vector $\blambda$, the same arguments discussed in Section~\ref{subsec:threshold} hold. 
		However, the statistics of the (Gaussian) randomly generated threshold vector $\boldsymbol{\lambda}$ change as follows:
		\begin{eqnarray}
			\mathbb{E}\left\{\boldsymbol{\lambda}\right\} &=& \mathbb{E} \{ \mathbf{y} \} = \mathbb{E}\left\{\alpha_0\right\} (\mathbf{s} \odot \mathbb{E} \{ \mathbf{p}(\nu) \}), \nonumber \\
			\text{Cov}(\boldsymbol{\lambda}) &=& \text{Cov} \left( \mathbf{y} \right) \\
			&=&\mathbb{E}\left\{|\alpha_0|^2 \right\} \, (\mathbf{s}  \mathbf{s}^H) \odot \mathbb{E} \{ \mathbf{p}(\nu) \mathbf{p}^H(\nu) \}  + \mathbf{R}. \nonumber
			\label{eq:lambda_char_Doppler}
		\end{eqnarray}

		For reader's convenience, the steps of the proposed approach for moving target radar parameter estimation are summarized in Algorithm~\ref{algorithm:optDoppler}.

		\begin{figure}
			\centering
			\includegraphics[width=\linewidth,draft=false]{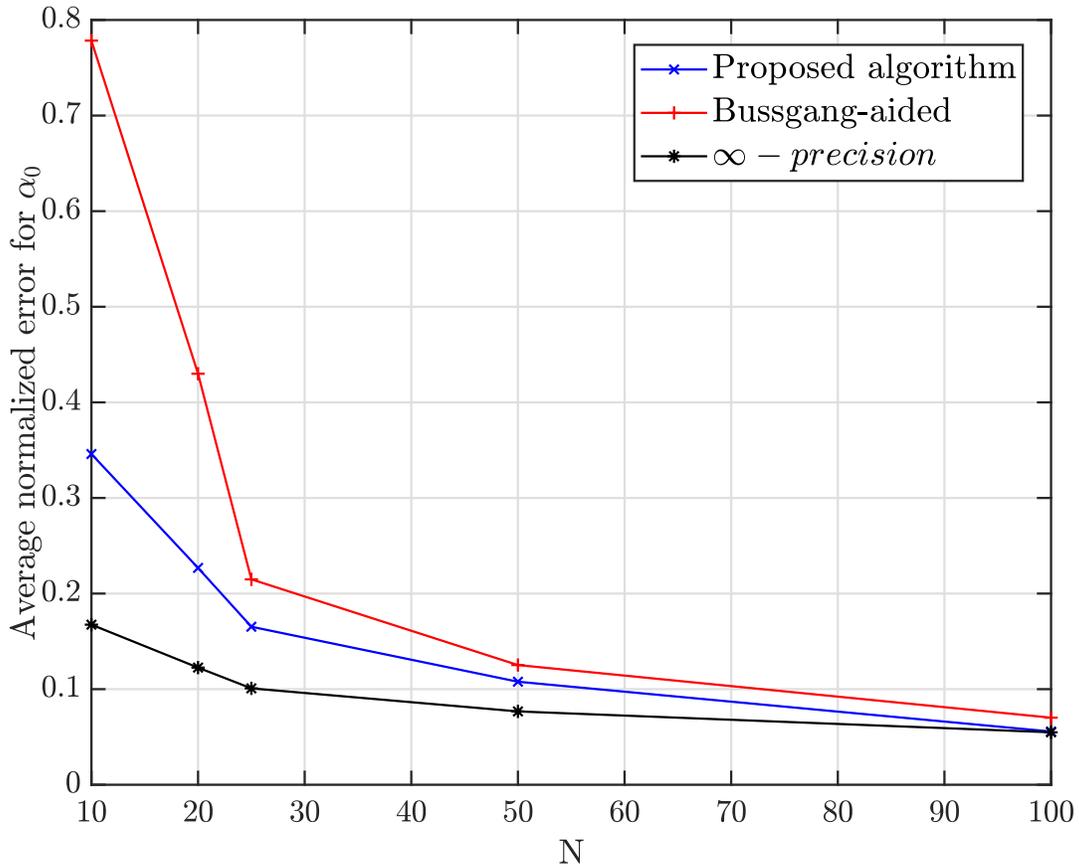}
			\caption{Average normalized estimation error of stationary target scattering coefficient $\alpha_0$, defined by the ratio $|\alpha_0 - \hat{\alpha}_0| / |\alpha_0|$, for different transmit sequence lengths $N \in \{10, 25, 50, 100\}$.}
			\label{fig:estimateVsN}	
		\end{figure}
		
		\begin{figure*}
			\centering
			\includegraphics[width=\textwidth,draft=false]{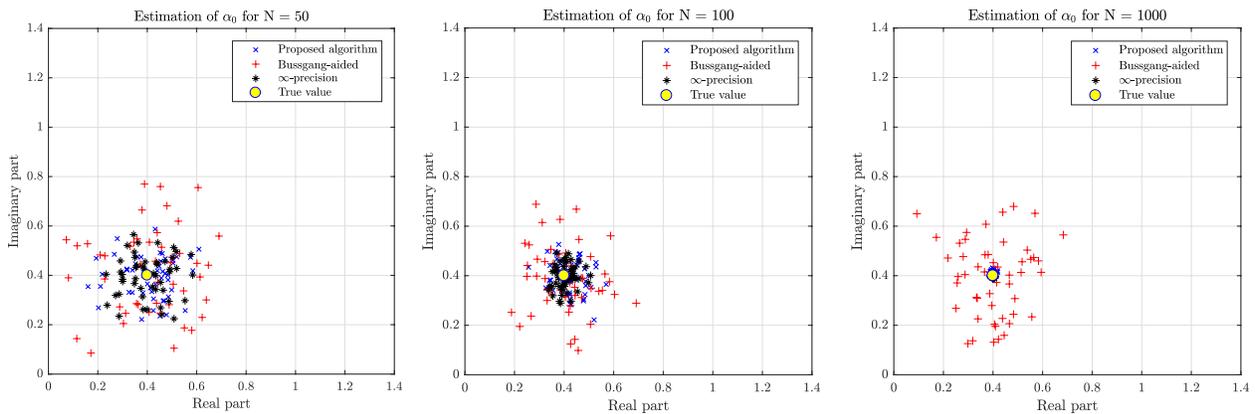}
			\caption{Comparison of stationary target scattering coefficient ($\alpha_0$) estimation performances for $N \in \{50, 100, 1000\}$. The results of estimation using the proposed algorithm, the Bussgang-aided approach, and the $\infty-precision$ case are shown on complex plane along with the true value of $\alpha_0=(0.5+j0.5)$.}
			\label{fig:estimateComplex}	
		\end{figure*}

\section{Extensions to Advanced Cases}
\label{sec:remarks}
	In this section, we study the extensions of the proposed method discussed in Section \ref{sec:proposedApproach} to different cases for the stationary target scenario.
	We further note that the same extensions can be applied to the moving targets case as well.
	\nocite{7676417}

	\subsection{Extension to Parallel One-Bit Comprators with Different Time-Varying Thresholds:}
		It can be noted that the problem formulation in \eqref{eq:opt_problem} can be extended to implementation of an array of $K$ number of one-bit comparators in parallel with different time-varying thresholds, denoted by $\boldsymbol{\lambda}^{(k)}, ~ k = 1, \cdots, K$.
		In this way, the optimization problem requires the recovered signal to comply with all the comparison information that are produced by the one-bit comparators.
		Thus, the constraints in \eqref{eq:opt_problem} can be updated as
		\begin{align}
			\boldsymbol{\Omega}_r^{(k)} \left( \mathbf{y}_r - \boldsymbol{\lambda}_r^{(k)} \right) \geq \mathbf{0}, \quad \forall~k \in \{1, \cdots, K\}, \nonumber \\
			\boldsymbol{\Omega}_i^{(k)} \left( \mathbf{y}_i - \boldsymbol{\lambda}_i^{(k)} \right) \geq \mathbf{0}, \quad \forall~k \in \{1, \cdots, K\},
		\end{align}
		where $\boldsymbol{\Omega}_r^{(k)} = \mathbf{Diag}(\boldsymbol{\lambda}_r^{(k)})$ and $\boldsymbol{\Omega}_i^{(k)} = \mathbf{Diag}(\boldsymbol{\lambda}_i^{(k)})$. 
	
	\subsection{Extension to $p$-Bit ADCs:}
		Another alternative way to glean more information from the received signal $\mathbf{y}$ is to use multi-bit ADCs.
		For a generic $p$-bit ADC, we have $(2^p-1)+2$ thresholds such that $\lambda_0 < \lambda_1 < \lambda_2 < \cdots < \lambda_{2^p-1} < \lambda_{2^p}$, where we define $\lambda_0 \triangleq -\infty$ and $\lambda_{2^p} \triangleq +\infty$ for ease of notation.
		Thus, each sample of the input signal can fall into any of the $2^p$ quantization regions, which further indicates that each input sampled data has to lie in an interval $\left[ \lambda_k, \lambda_{k+1} \right]$, for some $0 \leq k \leq (2^p-1), k \in \mathbb{N} \cup \{0\}$.
		Thus, if $q$ number of $p$-bit ADCs are used instead of one-bit comparators, each of the ADCs will have $(2^p-1)+2$ thresholds---leading to a total number of $q(2^p+1)$ thresholds.
	
		Observe that, the optimization problem in this case requires enforcing the following constraints,
		\begin{align*}
			[\mathbf{y}_r]_n &\in \left[ [\boldsymbol{\lambda}_r]_n^{(k_n)}, [\boldsymbol{\lambda}_r]_n^{(k_n+1)} \right], \\
			[\mathbf{y}_i]_m &\in \left[ [\boldsymbol{\lambda}_i]_m^{(k_m)}, [\boldsymbol{\lambda}_i]_m^{(k_m+1)} \right],
		\end{align*} 
		for all $n,m \in \{1,\cdots,N\}$ and for integers $k_n$ and $k_m$ provided by the $p$-bit ADCs, where $[\boldsymbol{\lambda}_r]_n^{(k_n)}$ and $[\boldsymbol{\lambda}_i]_m^{(k_m)}$ denote the $k_n$-th and $k_m$-th components of $[\boldsymbol{\lambda}_r]_n$ and $[\boldsymbol{\lambda}_i]_m$, respectively.
		Let, $\boldsymbol{\lambda}_r^{lower} \triangleq  \left[ [\boldsymbol{\lambda}_r]_1^{(k_1)}, \cdots, [\boldsymbol{\lambda}_r]_N^{(k_N)} \right]^T$,  $\boldsymbol{\lambda}_r^{upper} \triangleq  \left[ [\boldsymbol{\lambda}_r]_1^{(k_1+1)}, \cdots, [\boldsymbol{\lambda}_r]_N^{(k_N+1)} \right]^T$, and define  $\boldsymbol{\lambda}_i^{lower}$ and $\boldsymbol{\lambda}_i^{upper}$ in a similar manner. 
		Then, the constraints of the optimization problem in \eqref{eq:opt_problem} can be updated, in this case, as
		\begin{align}
		+1 \cdot (\mathbf{y}_r - \boldsymbol{\lambda}_r^{lower}) &\geq \mathbf{0}, \nonumber \\
		+1 \cdot (\mathbf{y}_i - \boldsymbol{\lambda}_i^{lower}) &\geq \mathbf{0}, \nonumber \\
		-1 \cdot (\mathbf{y}_r - \boldsymbol{\lambda}_r^{upper}) &\geq \mathbf{0}, \nonumber \\
		-1 \cdot (\mathbf{y}_i - \boldsymbol{\lambda}_i^{upper}) &\geq \mathbf{0}.
		\end{align}

	\subsection{Transition to Non-Negative Least-Squares (NNLS):} 
		It is  worth noting that the optimization problem in \eqref{eq:opt_problem} can easily be translated into a NNLS optimization problem.
		This can be achieved by changing variables such that $\tilde{\mathbf{y}}_r \triangleq \boldsymbol{\Omega}_r \left( \mathbf{y}_r - \boldsymbol{\lambda}_r \right)$ and $\tilde{\mathbf{y}}_i \triangleq \boldsymbol{\Omega}_i \left( \mathbf{y}_i - \boldsymbol{\lambda}_i \right)$, adding up to $\tilde{\mathbf{y}} \triangleq \tilde{\mathbf{y}}_r+j\tilde{\mathbf{y}}_i $.
		As a result, fast NNLS approaches can be exploited to expedite the recovery process \cite{bro1997fast}.

\section{Numerical Results}
\label{sec:numericalResults}
	In this section, we delve into examining the performance of the proposed target parameter estimation methods.
	The estimation error of our proposed approaches are compared with that of the Bussgang-aided approach of Section~\ref{sec:BussgangApproach}, and estimation using un-quantized received signal, denoted by $\infty-precision$.
	We first consider the case of stationary target and employ the approach discussed in Section~\ref{sec:proposedApproach} and then move on to the case of moving targets discussed in Section~\ref{sec:doppler}.

	\subsection{Stationary Targets}
	\label{sec:numericalResultsStationary}
		For the simulations, we assume that the noise is additive, white, and Gaussian with a variance of $0.1$, the average clutter power $\beta$ is $0.1$, and that the transmit sequence is generated using the method in \cite{soltanalian2013joint} with a peak-to-average power ratio of $1$.
		The results in all cases are averaged over 100 runs of the algorithms unless mentioned otherwise.

		Let $\hat{\alpha}_0$ denote the estimate of $\alpha_0$, and further define the normalized estimation error as $|\alpha_0 - \hat{\alpha}_0| / |\alpha_0|$.
		In Fig.~\ref{fig:estimateVsN}, the normalized estimation error obtained via a Monte-Carlo trial with randomly generated ground truths for $\alpha_0$ is plotted against the transmit sequence length $N$ for the proposed algorithm, the Bussgang-aided approach of Section~\ref{sec:BussgangApproach}, and the $\infty-precision$ case.
		In addition, for visualization purpose, Fig.~\ref{fig:estimateComplex} shows the results of the estimations, in a Monte-Carlo trial for $\alpha_0 = (0.5+j0.5)$ for $N \in \{50, 100, 1000\}$ on the complex plane.
		It can be seen from the both figures that the estimate of the proposed algorithm approaches that of the $\infty-precision$ as $N$ grows large.
		This is expected because when $N$ grows large, the number of comparisons grows large at the same rate revealing the true nature of the un-quantized data.
		From an information-theoretic point of view, this translates to more available information on the received signal through its one-bit samples that contribute to amelioration of the scattering coefficient recovery.
		Consequently, the estimation performance of all three approaches are enhanced with an increase of $N$, as is apparent in both figures.
		
\begin{figure}
			\centering
			\includegraphics[width=\linewidth,draft=false]{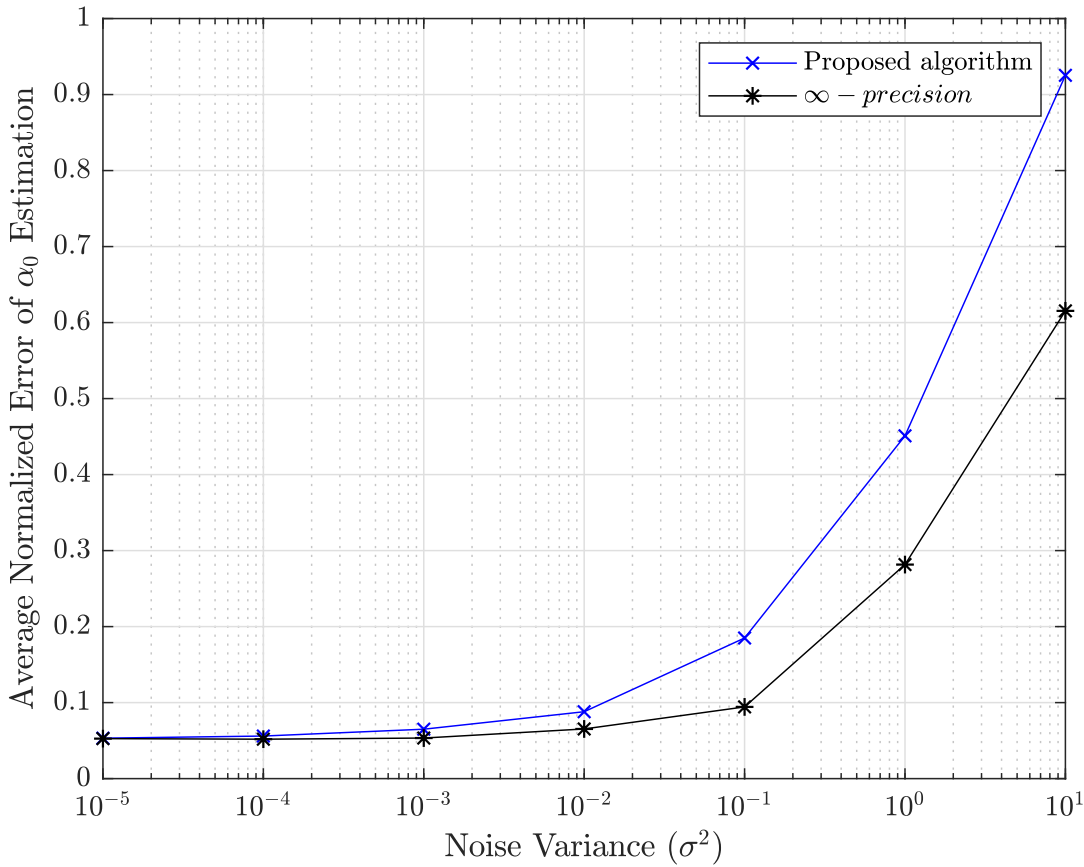}
			\caption{Average normalized estimation error of stationary target scattering coefficient $\alpha_0$, defined by the ratio $|\alpha_0 - \hat{\alpha}_0| / |\alpha_0|$, for different noise variances $\sigma^2 \in \{10^{-5}, 10^{-4}, 10^{-3}, 10^{-2}, 10^{-1}, 1, 10\}$.}
			\label{fig:noiseStationary}	
		\end{figure}

		Fig. \ref{fig:noiseStationary} shows the performance of the proposed algorithm in the presence of different noise power levels for the stationary target, and compares it with that of $\infty-precision$ case.
		For this experiment, we keep $N=25$, and again assume that the noise is additive, white, and Gaussian with variance $\sigma^2 \in \{10^{-5}, 10^{-4}, 10^{-3}, 10^{-2}, 10^{-1}, 1, 10\}$.
		As, it can be seen from Fig. \ref{fig:noiseStationary}, the average normalized error of estimating $\alpha_0$ remains very low for $\sigma^2 < 0.1$, however the performance of the algorithm decreases rapidly after $\sigma^2=1$.
		
	\subsection{Moving Targets}
	\label{sec:numericalResultsMoving}
		Herein we present the simulation results for radar parameter estimation in the case of moving targets.
		Similar to the stationary target scenario,  we assume that the noise is additive, white, and Gaussian with a variance of $0.1$, and that the transmit sequence is generated using the method in \cite{soltanalian2013joint} with a peak-to-average power ratio of $1$.
		The number of interfering range rings $N_c$ and number of azimuth sectors $L$ are set to $2$ and $10$, respectively. 
		Additionally, the normalized Doppler shifts of the adjacent range-azimuth cells are assumed to be uniformly distributed over the interval $\Omega_c = [-.1, .1]$; see \cite{skolnik1970radar} for further details. 
		
		\begin{figure*}
			\begin{minipage}[b]{0.48\linewidth}
				\centering
				\centerline{\includegraphics[width=0.38\textheight,draft=false]{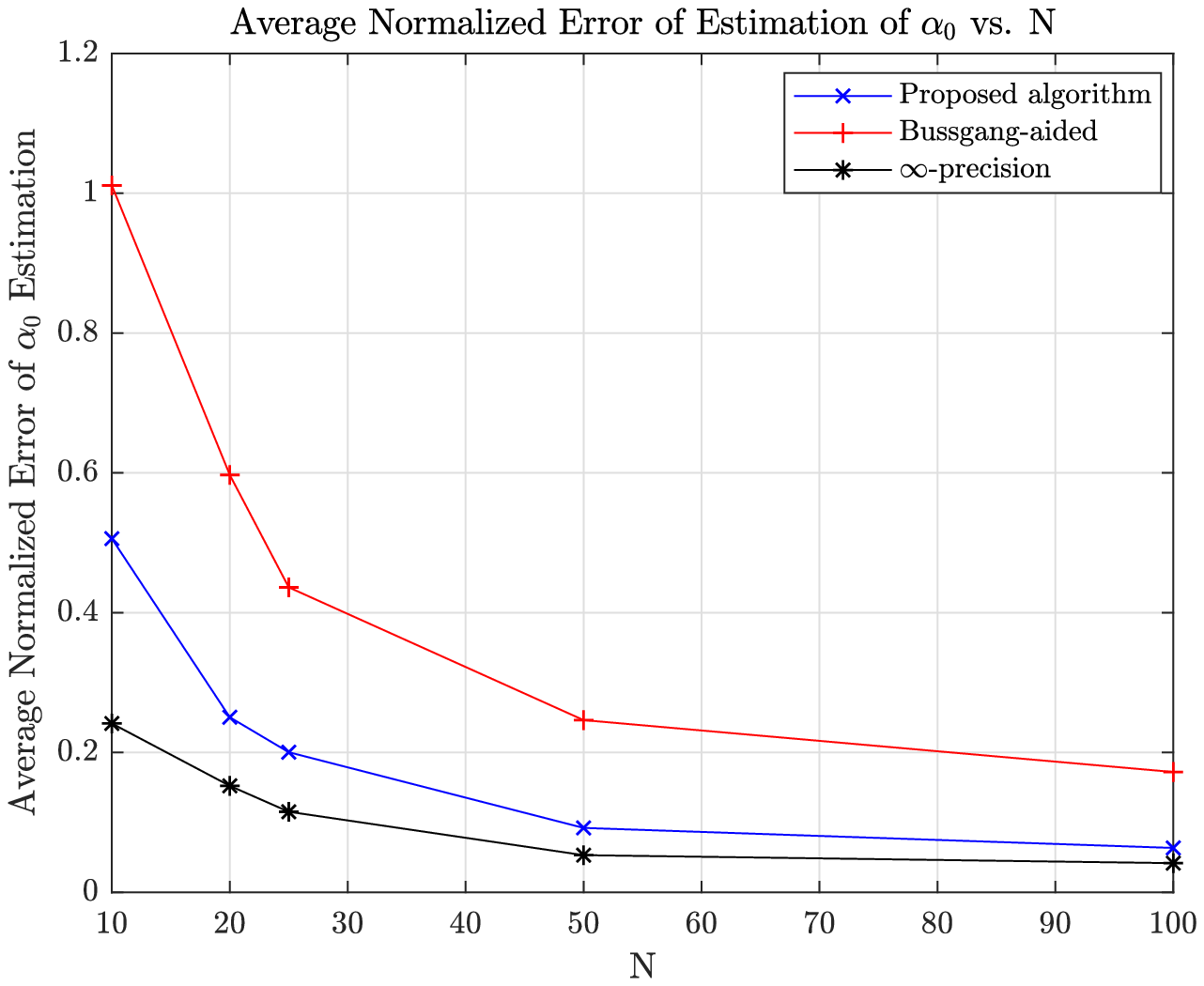}}  
				\centerline{(a)}\medskip
			\end{minipage}
			\hfill
			\begin{minipage}[b]{0.48\linewidth}
				\centering
				\centerline{\includegraphics[width=0.38\textheight,draft=false]{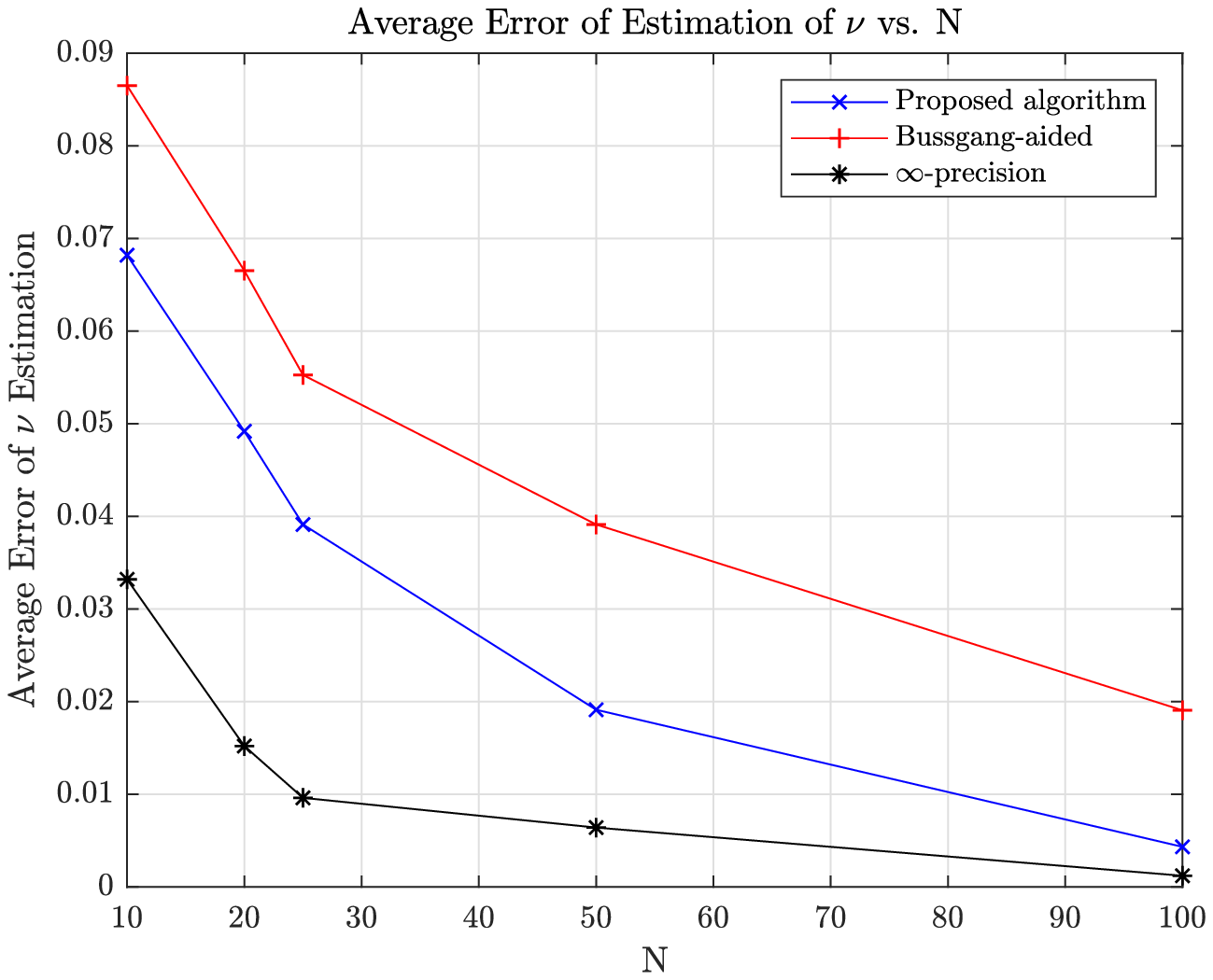}} 
				\centerline{(b)}\medskip
			\end{minipage}
			\caption{Performance comparison of moving target parameter estimation using the proposed algorithm, the Bussgang-aided approach, and the $\infty-precision$ case: (a) average normalized error of estimating the backscattering coefficient, ($|\alpha_0 - \hat{\alpha}_0| / |\alpha_0|$), (b) average error of estimating the normalized Doppler shift $\nu$, for different transmit sequence lengths $N \in \{10, 20, 25, 50, 100\}$.}
			\label{fig:estimateVsNDoppler}
		\end{figure*}
		\begin{figure*}
			\centering
			\includegraphics[width=\linewidth]{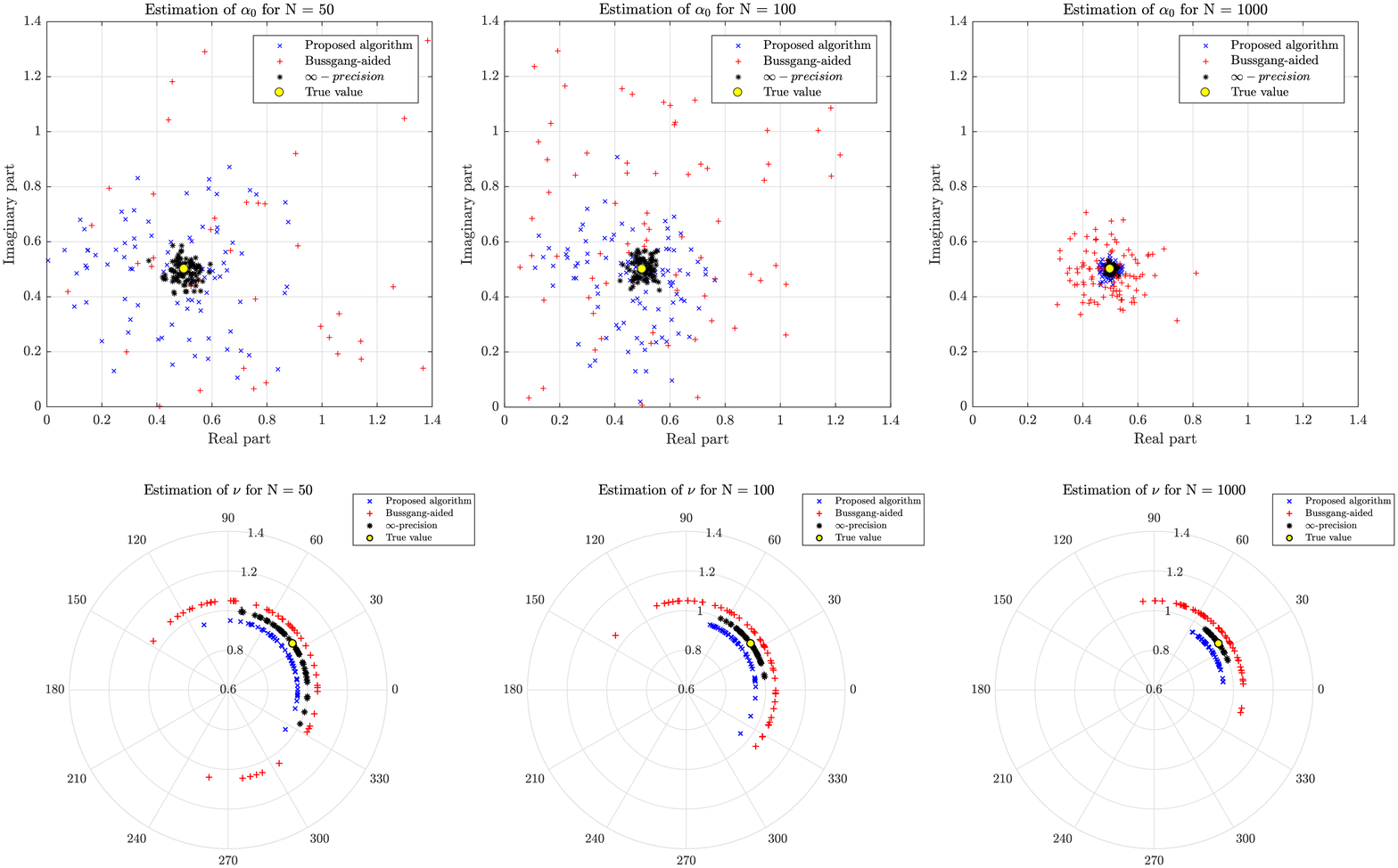}
			\caption{Performance comparison of moving target parameter estimation using the proposed approach, the Bussgang-aided approach, and the $\infty-precision$ case for $N \in \{50, 100, 1000\}$. The upper plots show the results of estimating $\alpha_0$ on the complex plane, while the lower plots show the results of estimating $\nu$ on the polar plane, where different radii are used for different approaches for visual clarity.}
			\label{fig:estimateComplexDoppler}	
		\end{figure*}
	
		The estimation performance of different approaches in moving target scenarios is examined via a Monte-Carlo trial with randomly generated ground truths for target parameters and presented in Fig.~\ref{fig:estimateVsNDoppler}.
		More precisely, the normalized estimation error of the proposed approach in estimating the backscattering coefficient $\alpha_0$ in case of a moving target, as well as the outcomes of the Bussgang-aided approach (modified for moving targets), and the $\infty-precision$ case are shown in Fig.~\ref{fig:estimateVsNDoppler}(a) while the errors for estimating the normalized Doppler shift $\nu$ are depicted in Fig.~\ref{fig:estimateVsNDoppler}(b).
		Furthermore, as in the case of a stationary target, Fig.~\ref{fig:estimateComplexDoppler} plots the radar parameter estimates for the case of a moving target through a Monte-Carlo trial. The upper plots in Fig.~\ref{fig:estimateComplexDoppler} show the results of estimating the backscattering coefficient $\alpha_0$, along with its true value, for $N \in \{50, 100, 1000\}$ on the complex plane.
		On the other hand, the lower plots in Fig.~\ref{fig:estimateComplexDoppler} show the estimates of the normalized Doppler shift on the polar plane.
		The result of estimation for different approaches are shown on circles with slightly different radii for the sake of clarity.

		As in the case of stationary targets, it can be observed from the Fig. \ref{fig:estimateVsNDoppler} that estimates $\alpha_0$ and $\nu$ become more precise as $N$ grows larger.
		In fact, as $N$ increases, the estimates of radar parameters obtained by the proposed approach get closer to that of the $\infty-precision$ case.
		Further note that in order to have the same performance in estimation of the parameters of a moving target, the proposed algorithm requires more samples than the stationary target case, as can be verified through Figs. \ref{fig:estimateVsN} and \ref{fig:estimateVsNDoppler} as anticipated.
		
\begin{figure*}[t]
			\begin{minipage}[b]{0.48\linewidth}
				\centering
				\centerline{\includegraphics[width=0.4\textheight,draft=false]{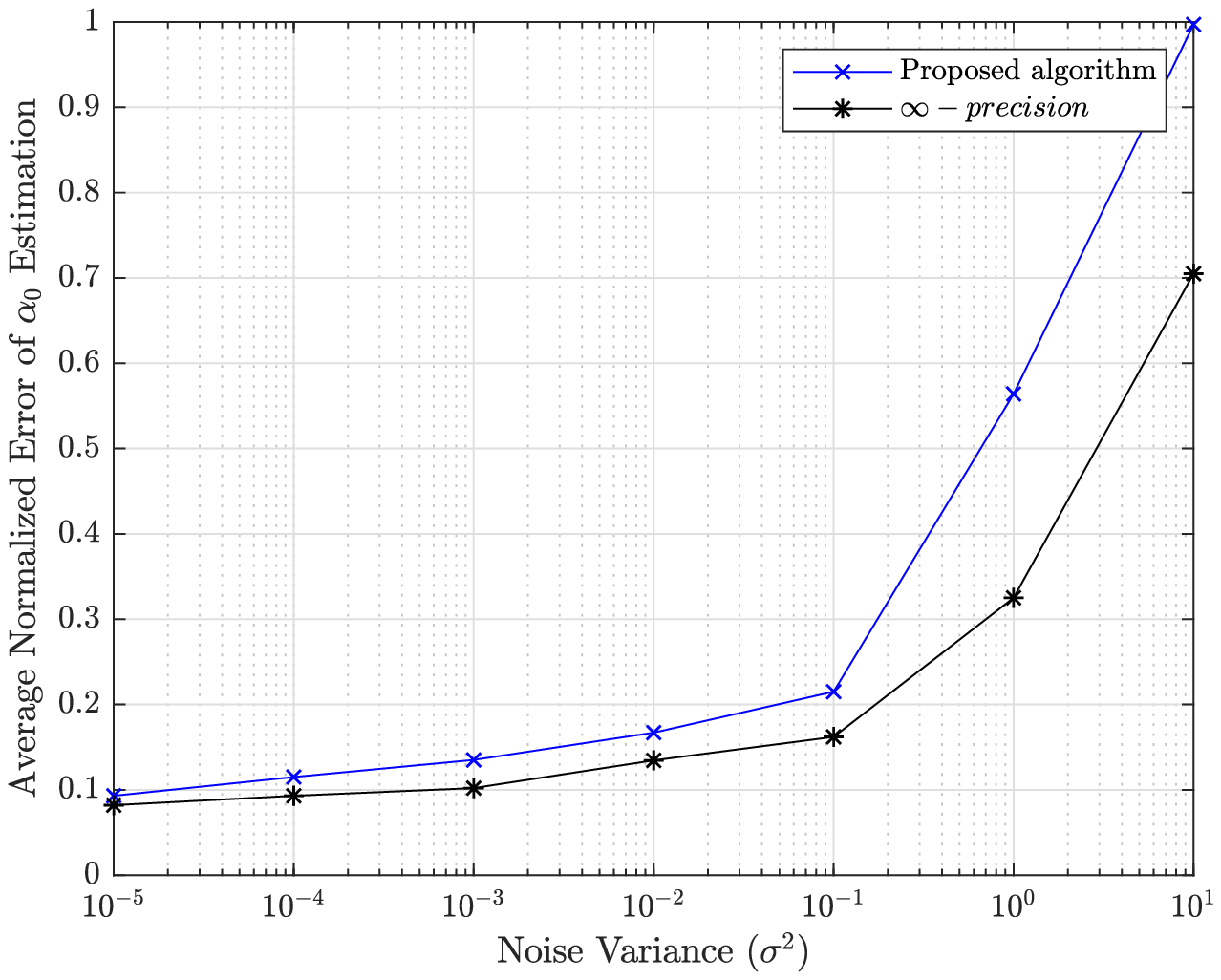}}  
				\centerline{(a)}\medskip
			\end{minipage}
			\hfill
			\begin{minipage}[b]{0.48\linewidth}
				\centering
				\centerline{\includegraphics[width=0.4\textheight,draft=false]{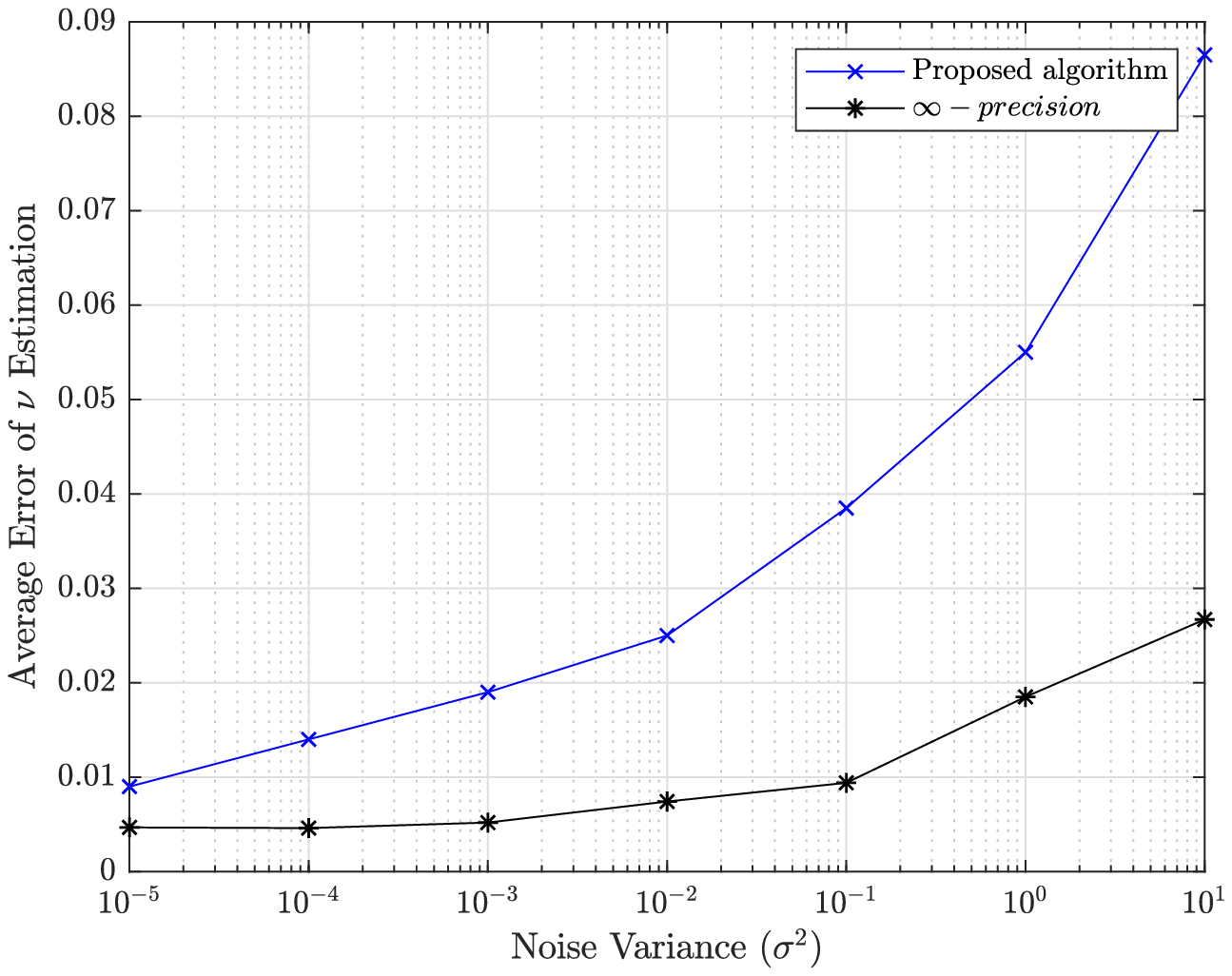}} 
				\centerline{(b)}\medskip
			\end{minipage}
			\caption{Performance comparison of moving target parameter estimation using the proposed algorithm, and the $\infty-precision$ case: (a) average normalized error of estimating the backscattering coefficient, ($|\alpha_0 - \hat{\alpha}_0| / |\alpha_0|$), (b) average error of estimating the normalized Doppler shift $\nu$, for different noise variances $\sigma^2 \in \{10^{-5}, 10^{-4}, 10^{-3}, 10^{-2}, 10^{-1}, 1, 10\}$.}
			\label{fig:noiseMoving}
		\end{figure*}

		Finally, Fig. \ref{fig:noiseMoving} demonstrates the performance of the proposed algorithm in the presence of different noise power levels for the moving targets, and compares it with that of $\infty-precision$ case.
		For this experiment, we again use the same settings as used for stationary case., i.e. $N=25$, and the noise is additive, white, and Gaussian with variance $\sigma^2 \in \{10^{-5}, 10^{-4}, 10^{-3}, 10^{-2}, 10^{-1}, 1, 10\}$.
		The average normalized error of estimating $\alpha_0$ is shown in Fig. \ref{fig:noiseMoving}(a) while the errors for estimating the normalized Doppler shift $\nu$ is depicted in Fig. \ref{fig:noiseMoving}(b).
		Similar to the case of stationary targets, it can be seen from Fig. \ref{fig:noiseMoving} that the average error of estimating both $\alpha_0$ and $\nu$ stay very low for $\sigma^2 < 0.1$, however their performances degrade after $\sigma^2=1$, even for $\infty-precision$ case.
	
\section{Concluding Remarks}
\label{sec:conclusion}
	High-resolution sampling with conventional analog-to-digital-converters (ADCs) can be very costly and energy-consuming for many modern applications.
	This is further accentuated as recent applications, including those in sensing and radar signal processing, show a growing appetite in even larger than usual sampling rates--- thus making the mainstream ADCs a rather unsuitable choice.
	To overcome these shortcomings, it was shown that in lieu of using the conventional ADCs in radar parameters estimation, one can use inexpensive comparators with time-varying thresholds and solve an optimization problem to recover the target parameters with satisfactory performance.
	This is very beneficial at high frequencies as it is both practical and economical, while it can also pave the way for future applications to sample at much higher rates.
	Finally, simulation results were presented that verified the efficiency of one-bit target parameter estimation for both stationary and moving targets, especially as the length of the transmit sequence $N$ grows large.

\bibliographystyle{IEEEbib}
\bibliography{refs}

\begin{thebibliography}{10}

\bibitem{ameri2018one}
A.~{Ameri}, J.~{Li}, and M.~{Soltanalian},
\newblock ``One-bit radar processing and estimation with time-varying sampling
  thresholds,''
\newblock in {\em 2018 IEEE 10th Sensor Array and Multichannel Signal
  Processing Workshop (SAM)}, July 2018, pp. 208--212.

\bibitem{stepanenko1975radar}
V.~D. Stepanenko,
\newblock ``Radar in meteorology,''
\newblock Tech. {R}ep., Foreign technology div Wright-Patterson AFB Ohio, 1975.

\bibitem{Radarmeteorology}
K.~Browning,
\newblock ``Uses of radar in meteorology,''
\newblock {\em Contemporary Physics}, vol. 27, no. 6, pp. 499--517, 1986.

\bibitem{482830}
S.~Bussolari and J.~Bernays,
\newblock ``Mode {S} data link applications for general aviation,''
\newblock in {\em Proceedings of 14th Digital Avionics Systems Conference}, Nov
  1995, pp. 199--206.

\bibitem{nolan2010fundamentals}
M.~Nolan,
\newblock {\em Fundamentals of air traffic control},
\newblock Cengage learning, 2010.

\bibitem{ihn2008pitch}
J.~B. Ihn and F.~K. Chang,
\newblock ``Pitch-catch active sensing methods in structural health monitoring
  for aircraft structures,''
\newblock {\em Structural Health Monitoring}, vol. 7, no. 1, pp. 5--19, 2008.

\bibitem{lynch2004design}
J.~P. Lynch, A.~Sundararajan, K.~H. Law, H.~Sohn, and C.~R. Farrar,
\newblock ``Design of a wireless active sensing unit for structural health
  monitoring,''
\newblock in {\em Health Monitoring and Smart Nondestructive Evaluation of
  Structural and Biological Systems III}. International Society for Optics and
  Photonics, 2004, vol. 5394, pp. 157--169.

\bibitem{soumekh1999synthetic}
M.~Soumekh,
\newblock {\em Synthetic aperture radar signal processing}, vol.~7,
\newblock New York: Wiley, 1999.

\bibitem{curlander1991synthetic}
J.~C. Curlander and R.~N. McDonough,
\newblock {\em Synthetic aperture radar}, vol. 396,
\newblock John Wiley \& Sons New York, NY, USA, 1991.

\bibitem{heidemann2012underwater}
J.~Heidemann, M.~Stojanovic, and M.~Zorzi,
\newblock ``Underwater sensor networks: applications, advances and
  challenges,''
\newblock {\em Phil. Trans. R. Soc. A}, vol. 370, no. 1958, pp. 158--175, 2012.

\bibitem{farr2010integrated}
N.~Farr, A.~Bowen, J.~Ware, C.~Pontbriand, and M.~Tivey,
\newblock ``An integrated, underwater optical/acoustic communications system,''
\newblock in {\em OCEANS}. IEEE, 2010, pp. 1--6.

\bibitem{rummler1967technique}
W.~D. Rummler,
\newblock ``A technique for improving the clutter performance of coherent pulse
  train signals,''
\newblock {\em IEEE Transactions on Aerospace and Electronic Systems}, pp.
  898--906, 1967.

\bibitem{pillai1999optimum}
S.~U. Pillai, D.~C. Youla, H.~S. Oh, and J.~R. Guerci,
\newblock ``Optimum transmit-receiver design in the presence of
  signal-dependent interference and channel noise,''
\newblock in {\em Signals, Systems, and Computers, 1999. Conference Record of
  the Thirty-Third Asilomar Conference on}. IEEE, 1999, vol.~2, pp. 870--875.

\bibitem{delong1967design}
D.~DeLong and E.~Hofstetter,
\newblock ``On the design of optimum radar waveforms for clutter rejection,''
\newblock {\em IEEE Transactions on Information Theory}, vol. 13, no. 3, pp.
  454--463, 1967.

\bibitem{bell1993information}
M.~R. Bell,
\newblock ``Information theory and radar waveform design,''
\newblock {\em IEEE Transactions on Information Theory}, vol. 39, no. 5, pp.
  1578--1597, 1993.

\bibitem{kay2007optimal}
S.~Kay,
\newblock ``Optimal signal design for detection of {G}aussian point targets in
  stationary clutter/reverberation,''
\newblock {\em IEEE journal of selected topics in signal processing}, vol. 1,
  no. 1, pp. 31--41, 2007.

\bibitem{stoica2008transmit}
P.~Stoica, J.~Li, and M.~Xue,
\newblock ``Transmit codes and receive filters for radar,''
\newblock {\em IEEE Signal Processing Magazine}, vol. 25, no. 6, 2008.

\bibitem{stoica2012optimization}
P.~Stoica, H.~He, and J.~Li,
\newblock ``Optimization of the receive filter and transmit sequence for active
  sensing,''
\newblock {\em IEEE Transactions on Signal Processing}, vol. 60, no. 4, pp.
  1730--1740, 2012.

\bibitem{soltanalian2013joint}
M.~Soltanalian, B.~Tang, J.~Li, and P.~Stoica,
\newblock ``Joint design of the receive filter and transmit sequence for active
  sensing,''
\newblock {\em IEEE Signal Processing Letters}, vol. 20, no. 5, pp. 423--426,
  2013.

\bibitem{naghsh2014doppler}
M.~M. Naghsh, M.~Soltanalian, P.~Stoica, M.~Modarres-Hashemi, A.~De~Maio, and
  A.~Aubry,
\newblock ``A {Doppler} robust design of transmit sequence and receive filter
  in the presence of signal-dependent interference,''
\newblock {\em IEEE Transactions on Signal Processing}, vol. 62, no. 4, pp.
  772--785, 2014.

\bibitem{he2012waveform}
H.~He, J.~Li, and P.~Stoica,
\newblock {\em Waveform design for active sensing systems: a computational
  approach},
\newblock Cambridge University Press, 2012.

\bibitem{radarSignals}
N.~Levanon and E.~Mozeson,
\newblock {\em Radar signals},
\newblock John Wiley \& Sons, 2004.

\bibitem{woodward2014probability}
P.~M. Woodward,
\newblock {\em Probability and Information Theory, with Applications to Radar:
  International Series of Monographs on Electronics and Instrumentation},
  vol.~3,
\newblock Elsevier, 2014.

\bibitem{1057703}
S.~{Sussman},
\newblock ``Least-square synthesis of radar ambiguity functions,''
\newblock {\em IRE Transactions on Information Theory}, vol. 8, no. 3, pp.
  246--254, April 1962.

\bibitem{4103366}
J.~D. {Wolf}, G.~M. {Lee}, and C.~E. {Suyo},
\newblock ``Radar waveform synthesis by mean square optimization techniques,''
\newblock {\em IEEE Transactions on Aerospace and Electronic Systems}, vol.
  AES-5, no. 4, pp. 611--619, July 1969.

\bibitem{4749273}
P.~{Stoica}, H.~{He}, and J.~{Li},
\newblock ``New algorithms for designing unimodular sequences with good
  correlation properties,''
\newblock {\em IEEE Transactions on Signal Processing}, vol. 57, no. 4, pp.
  1415--1425, April 2009.

\bibitem{6142119}
M.~{Soltanalian} and P.~{Stoica},
\newblock ``Computational design of sequences with good correlation
  properties,''
\newblock {\em IEEE Transactions on Signal Processing}, vol. 60, no. 5, pp.
  2180--2193, May 2012.

\bibitem{8314765}
A.~{Bose} and M.~{Soltanalian},
\newblock ``Constructing binary sequences with good correlation properties: An
  efficient analytical-computational interplay,''
\newblock {\em IEEE Transactions on Signal Processing}, vol. 66, no. 11, pp.
  2998--3007, June 2018.

\bibitem{4644058}
P.~Stoica, J.~Li, and M.~Xue,
\newblock ``Transmit codes and receive filters for radar,''
\newblock {\em IEEE Signal Processing Magazine}, vol. 25, no. 6, pp. 94--109,
  Nov 2008.

\bibitem{1054205}
L.~{Spafford},
\newblock ``Optimum radar signal processing in clutter,''
\newblock {\em IEEE Transactions on Information Theory}, vol. 14, no. 5, pp.
  734--743, Sep. 1968.

\bibitem{gianelli2016one}
C.~Gianelli, L.~Xu, J.~Li, and P.~Stoica,
\newblock ``One-bit compressive sampling with time-varying thresholds for
  sparse parameter estimation,''
\newblock in {\em Sensor Array and Multichannel Signal Processing Workshop
  (SAM), 2016 IEEE}. IEEE, 2016, pp. 1--5.

\bibitem{sun2013wideband}
H.~Sun, A.~Nallanathan, C.~X. Wang, and Y.~Chen,
\newblock ``Wideband spectrum sensing for cognitive radio networks: a survey,''
\newblock {\em IEEE Wireless Communications}, vol. 20, no. 2, pp. 74--81, 2013.

\bibitem{lunden2015spectrum}
J.~Lunden, V.~Koivunen, and H.~V. Poor,
\newblock ``Spectrum exploration and exploitation for cognitive radio: Recent
  advances,''
\newblock {\em IEEE signal processing magazine}, vol. 32, no. 3, pp. 123--140,
  2015.

\bibitem{burke2009introduction}
B.~F. Burke and F.~Graham-Smith,
\newblock {\em An introduction to radio astronomy},
\newblock Cambridge University Press, 2009.

\bibitem{strohm2005development}
K.~M. Strohm, H.~L. Bloecher, R.~Schneider, and J.~Wenger,
\newblock ``Development of future short range radar technology,''
\newblock in {\em Radar Conference, 2005. EURAD 2005. European}. IEEE, 2005,
  pp. 165--168.

\bibitem{hasch2012millimeter}
J.~Hasch, E.~Topak, R.~Schnabel, T.~Zwick, R.~Weigel, and C.~Waldschmidt,
\newblock ``Millimeter-wave technology for automotive radar sensors in the 77
  {GHz} frequency band,''
\newblock {\em IEEE Transactions on Microwave Theory and Techniques}, vol. 60,
  no. 3, pp. 845--860, 2012.

\bibitem{8645383}
S.~{Khobahi} and M.~{Soltanalian},
\newblock ``Signal recovery from 1-bit quantized noisy samples via adaptive
  thresholding,''
\newblock in {\em 2018 52nd Asilomar Conference on Signals, Systems, and
  Computers}, Oct 2018, pp. 1757--1761.

\bibitem{ribeiro2006bandwidthi}
A.~Ribeiro and G.~B. Giannakis,
\newblock ``Bandwidth-constrained distributed estimation for wireless sensor
  networks-part i: {Gaussian} case,''
\newblock {\em IEEE transactions on signal processing}, vol. 54, no. 3, pp.
  1131--1143, 2006.

\bibitem{ribeiro2006bandwidthii}
A.~Ribeiro and G.~B. Giannakis,
\newblock ``Bandwidth-constrained distributed estimation for wireless sensor
  networks-part ii: unknown probability density function,''
\newblock {\em IEEE Transactions on Signal Processing}, vol. 54, no. 7, pp.
  2784--2796, 2006.

\bibitem{host2000effects}
A.~Host-Madsen and P.~Handel,
\newblock ``Effects of sampling and quantization on single-tone frequency
  estimation,''
\newblock {\em IEEE Transactions on Signal Processing}, vol. 48, no. 3, pp.
  650--662, 2000.

\bibitem{bar2002doa}
O.~Bar-Shalom and A.~J. Weiss,
\newblock ``{DOA} estimation using one-bit quantized measurements,''
\newblock {\em IEEE Transactions on Aerospace and Electronic Systems}, vol. 38,
  no. 3, pp. 868--884, 2002.

\bibitem{dabeer2006signal}
O.~Dabeer and A.~Karnik,
\newblock ``Signal parameter estimation using 1-bit dithered quantization,''
\newblock {\em IEEE Transactions on Information Theory}, vol. 52, no. 12, pp.
  5389--5405, 2006.

\bibitem{dabeer2008multivariate}
O.~Dabeer and E.~Masry,
\newblock ``Multivariate signal parameter estimation under dependent noise from
  1-bit dithered quantized data,''
\newblock {\em IEEE Transactions on Information Theory}, vol. 54, no. 4, pp.
  1637--1654, 2008.

\bibitem{8683876}
S.~{Khobahi}, N.~{Naimipour}, M.~{Soltanalian}, and Y.~C. {Eldar},
\newblock ``Deep signal recovery with one-bit quantization,''
\newblock in {\em 2019 IEEE International Conference on Acoustics, Speech and
  Signal Processing (ICASSP)}, May 2019, pp. 2987--2991.

\bibitem{7676417}
H.~{Hu}, M.~{Soltanalian}, P.~{Stoica}, and X.~{Zhu},
\newblock ``Locating the few: Sparsity-aware waveform design for active
  radar,''
\newblock {\em IEEE Transactions on Signal Processing}, vol. 65, no. 3, pp.
  651--662, Feb 2017.

\bibitem{comp_sense_1}
P.~Boufounos and R.~Baraniuk,
\newblock ``1-bit compressive sensing,''
\newblock in {\em 42nd Annual Conference on Information Sciences and Systems},
  Mar 2008, pp. 16--21.

\bibitem{comp_sense_3}
K.~Knudson, R.~Saab, and R.~Ward,
\newblock ``One-bit compressive sensing with norm estimation,''
\newblock {\em IEEE Transactions on Information Theory}, vol. 62, no. 5, pp.
  2748--2758, May 2016.

\bibitem{comp_sense_4}
A.~Zymnis, S.~Boyd, and E.~Candes,
\newblock ``Compressed sensing with quantized measurements,''
\newblock {\em IEEE Signal Processing Letters}, vol. 17, no. 2, pp. 149--152,
  Feb 2010.

\bibitem{comp_sense_5}
Y.~Plan and R.~Vershynin,
\newblock ``Robust 1-bit compressed sensing and sparse logistic regression: A
  convex programming approach,''
\newblock {\em IEEE Transactions on Information Theory}, vol. 59, no. 1, pp.
  482--494, Jan 2013.

\bibitem{dong2015map}
X.~Dong, Y.~Zhang, and {others},
\newblock ``A map approach for 1-bit compressive sensing in synthetic aperture
  radar imaging.,''
\newblock {\em IEEE Geosci. Remote Sensing Lett.}, vol. 12, no. 6, pp.
  1237--1241, 2015.

\bibitem{masry1980reconstruction}
E.~Masry,
\newblock ``The reconstruction of analog signals from the sign of their noisy
  samples,''
\newblock Tech. {R}ep., California university of San Diego, La Jolla, dept of
  electrical engineering and computer sciences, 1980.

\bibitem{cvetkovic2000single}
Z.~Cvetkovic and I.~Daubechies,
\newblock ``Single-bit oversampled {A/D} conversion with exponential accuracy
  in the bit-rate,''
\newblock in {\em Data Compression Conference, 2000. Proceedings. DCC 2000}.
  IEEE, 2000, pp. 343--352.

\bibitem{1642573}
S.~Blunt and K.~Gerlach,
\newblock ``Adaptive pulse compression via {MMSE} estimation,''
\newblock {\em IEEE Transactions on Aerospace and Electronic Systems}, vol. 42,
  no. 2, pp. 572--584, Apr 2006.

\bibitem{cognitiveRadar}
S.~Haykin,
\newblock ``Cognitive radar: a way of the future,''
\newblock {\em IEEE Signal Processing Magazine}, vol. 23, no. 1, pp. 30--40,
  Jan 2006.

\bibitem{bussgang1952crosscorrelation}
J.~J. Bussgang,
\newblock ``Crosscorrelation functions of amplitude-distorted {Gaussian}
  signals,''
\newblock 1952.

\bibitem{van1966spectrum}
J.~H. Van~Vleck and D.~Middleton,
\newblock ``The spectrum of clipped noise,''
\newblock {\em Proceedings of the IEEE}, vol. 54, no. 1, pp. 2--19, 1966.

\bibitem{liu2017one}
C.~L. Liu and P.~P. Vaidyanathan,
\newblock ``One-bit sparse array {DOA} estimation,''
\newblock in {\em Acoustics, Speech and Signal Processing (ICASSP), 2017 IEEE
  International Conference on}. IEEE, 2017, pp. 3126--3130.

\bibitem{6407265}
A.~Aubry, A.~De Maio, M.~Piezzo, A.~Farina, and M.~Wicks,
\newblock ``Cognitive design of the receive filter and transmitted phase code
  in reverberating environment,''
\newblock {\em IET Radar, Sonar Navigation}, vol. 6, no. 9, pp. 822--833, Dec
  2012.

\bibitem{gini2012waveform}
F.~Gini, A.~De~Maio, and L.~Patton,
\newblock {\em Waveform design and diversity for advanced radar systems},
\newblock Institution of engineering and technology London, 2012.

\bibitem{6404093}
A.~Aubry, A.~De Maio, A.~Farina, and M.~Wicks,
\newblock ``Knowledge-aided (potentially cognitive) transmit signal and receive
  filter design in signal-dependent clutter,''
\newblock {\em IEEE Transactions on Aerospace and Electronic Systems}, vol. 49,
  no. 1, pp. 93--117, Jan 2013.

\bibitem{riverSurfaceCurrents}
C.~J. Wang, B.~Y. Wen, Z.~G. Ma, W.~D. Yan, and X.~J. Huang,
\newblock ``Measurement of river surface currents with uhf fmcw radar
  systems,''
\newblock {\em Journal of Electromagnetic Waves and Applications}, vol. 21, no.
  3, pp. 375--386, 2007.

\bibitem{923012}
M.~{Kondo}, K.~{Kawai}, H.~{Hirano}, and T.~{Fujisaka},
\newblock ``Ocean wave observation by cw mm-wave radar with narrow beam,''
\newblock in {\em Proceedings of the 2001 IEEE Radar Conference}, May 2001, pp.
  398--403.

\bibitem{stoica2005spectral}
P.~Stoica, R.~L. Moses, and {others},
\newblock ``Spectral analysis of signals,''
\newblock 2005.

\bibitem{bro1997fast}
R.~Bro and S.~De~Jong,
\newblock ``A fast non-negativity-constrained least squares algorithm,''
\newblock {\em Journal of chemometrics}, vol. 11, no. 5, pp. 393--401, 1997.

\bibitem{skolnik1970radar}
M.~I. Skolnik,
\newblock ``Radar handbook,''
\newblock 1970.

\end{thebibliography}

\end{document}